\documentclass[twocolumn]{aastex61}
\pdfoutput=1 
\usepackage{amsmath,amstext}
\usepackage[T1]{fontenc}
\usepackage{apjfonts} 
\usepackage[figure,figure*]{hypcap}
\usepackage{url} 
\usepackage{xcolor}

\shorttitle{Elemental Abundances of KOIs in APOGEE. I.}
\shortauthors{Wilson et al.}

\renewcommand{\micron}{$\mu$m}
\newcommand{\teff}{$T_\mathrm{eff}$ }
\newcommand{\pcrit}{$P_\mathrm{crit}$}

\begin{document}

\title{Elemental Abundances of Kepler Objects of Interest in APOGEE. I. Two Distinct Orbital Period Regimes Inferred from Host Star Iron Abundances}

\author{Robert F. Wilson}
\affiliation{Department of Astronomy, University of Virginia, Charlottesville, VA 22904-4325, USA}

\author{Johanna Teske}
\affiliation{Observatories of the Carnegie Institution for Science, 813 Santa Barbara St., Pasadena, CA 91101}
\affiliation{Carnegie Origins Fellow, jointly appointed by Carnegie DTM and Observatories} 

\author{Steven R. Majewski}
\affiliation{Department of Astronomy, University of Virginia, Charlottesville, VA 22904-4325, USA}

\author{Katia Cunha}
\affiliation{Observat\'orio Nacional, Rua General Jos\'e Cristino, 77, 20921-400 S\~ao Crist\'ov\~ao, Rio de Janeiro, RJ, Brazil}

\affiliation{Steward Observatory, University of Arizona, 933 North Cherry Avenue, Tucson, AZ 85721-0065, USA}

\author{Verne Smith}
\affiliation{National Optical Astronomy Observatory, 950 North Cherry Avenue, Tucson, AZ 85719, USA}

\author{Diogo Souto}
\affiliation{Observat\'orio Nacional, Rua General Jos\'e Cristino, 77, 20921-400 S\~ao Crist\'ov\~ao, Rio de Janeiro, RJ, Brazil}

\author{Chad Bender}
\affiliation{Steward Observatory, University of Arizona, 933 North Cherry Avenue, Tucson, AZ 85721-0065, USA}

\author{Suvrath Mahadevan}
\affiliation{Department of Astronomy \& Astrophysics, Pennsylvania State University, University Park, PA 16802, USA	}

\author{Nicholas Troup}
\affiliation{Department of Astronomy, University of Virginia, Charlottesville, VA 22904-4325, USA}
\affiliation{Department of Physics, Salisbury University, Salisbury, MD 21801-6860, USA}

\author{Carlos Allende Prieto}
\affiliation{Instituto de Astrof\'isica de Canarias, E-38205 La Laguna, Tenerife, Spain}
\affiliation{Departamento de Astrof\'isica, Universidad de La Laguna, E-38206 La Laguna, Tenerife, Spain}

\author{Keivan G. Stassun}
\affiliation{Department of Physics and Astronomy, Vanderbilt University, VU Station 1807, Nashville, TN 37235, USA}

\author{Michael F. Skrutskie}
\affiliation{Department of Astronomy, University of Virginia, Charlottesville, VA 22904-4325, USA}

\author{Andr\'{e}s Almeida}
\affiliation{Instituto de Investigaci\'{o}n Multidisciplinario en Ciencia y Tecnolog\'{i}a, Universidad de La Serena, Benavente 980, La Serena, Chile}

\author{D. A. Garc\'{i}a-Hern\'{a}ndez}
\affiliation{Instituto de Astrof\'isica de Canarias, E-38205 La Laguna, Tenerife, Spain}
\affiliation{Departamento de Astrof\'isica, Universidad de La Laguna, E-38206 La Laguna, Tenerife, Spain}

\author{Olga Zamora}
\affiliation{Instituto de Astrof\'isica de Canarias, E-38205 La Laguna, Tenerife, Spain}
\affiliation{Departamento de Astrof\'isica, Universidad de La Laguna, E-38206 La Laguna, Tenerife, Spain}

\author{Jonathan Brinkmann}
\affiliation{Apache Point Observatory, P.O. Box 59, Sunspot, NM 88349}

\begin{abstract}

The Apache Point Observatory Galactic Evolution Experiment (APOGEE) has observed $\sim$600 transiting exoplanets and exoplanet candidates from \textit{Kepler} (Kepler Objects of Interest, KOIs), most with $\geq$18 epochs. The combined multi-epoch spectra are of high signal-to-noise (typically $\geq$100) and yield precise stellar parameters and chemical abundances. We first confirm the ability of the APOGEE abundance pipeline, ASPCAP, to derive reliable [Fe/H] and effective temperatures for FGK dwarf stars --- the primary \textit{Kepler} host stellar type --- by comparing the ASPCAP-derived stellar parameters to those from independent high-resolution spectroscopic characterizations for 221 dwarf stars in the literature. With a sample of 282 close-in ($P<100$ days) KOIs observed in the APOGEE KOI goal program, we find a correlation between orbital period and host star [Fe/H] characterized by a critical period,  \pcrit = $8.3^{+0.1}_{-4.1}$ days, below which small exoplanets orbit statistically more metal-enriched host stars. This effect may trace a metallicity dependence of the protoplanetary disk inner-radius at the time of planet formation or may be a result of rocky planet ingestion driven by inward planetary migration. We also consider that this may trace a metallicity dependence of the dust sublimation radius, but find no statistically significant correlation with host \teff and orbital period to support such a claim.

\end{abstract}


%
%
\section{Introduction}

With the advent of the \textit{Kepler} mission \citep{koch2010,borucki2010,borucki2016}, statistical studies of exoplanets, particularly small planets ($R_p \lesssim 4 R_{\oplus}$), have become possible. While a key finding of such studies is that small planets are common in the Galaxy in general \citep[e.g.,][]{howard2012,dressing&charbonneau2013,petigura2013,batalha2014,burke2015,silburt2015}, distinguishing the characteristics of these planets and how they 
may relate
to the properties of their host stars is 
of interest from formation 
and detection perspectives. From population studies of larger planets detected by the radial velocity method, it was clear early on that host star metallicity\footnote{Usually parameterized by the number density of iron nuclei in a star's photosphere relative to the amount of hydrogen, normalized to these values in the Sun: [Fe/H], where [X/H]=log(N$_{\rm{X}}$/N$_{\rm{H}}$) - log (N$_{\rm{X}}$/N$_{\rm{H}}$)$_{\rm{solar}}$} was related to the frequency at which these planets form \citep{gonzalez1998,heiter2003,santos2004,valenti&fischer2005}, a trend that appears to decrease in strength with decreasing planet mass and/or radius \citep[e.g.,][]{sousa2008,ghezzi2010,schlaufman&laughlin2011, buchhave2012, wang&fischer2015, buchhave&latham2015}. Now, the most prevalent explanation of this trend is that it is evidence of the core accretion method of planet formation \citep[e.g.,][]{rice&armitage2003,ida&lin2004,alibert2011,mordasini2012}, and that host star metallicity is a proxy for the surface density of the solid material in a protoplanetary disk; higher solid surface densities facilitate the faster growth of the solid cores of larger planets, giving them more time to accrete gaseous envelopes.

In addition to the trend between host star [Fe/H] and the frequency of different types of planets, 
other relationships between
stellar metallicity and planet properties have also come to light. For instance, \cite{dawson&murrayclay2013} used the evidence that giant planets orbiting [Fe/H]$<$0 stars generally have lower eccentricity orbits to suggest that planet-planet scattering is the dominant mechanism for inward migration of giant planets, since higher [Fe/H] systems are more likely to form multiple, closely spaced giant planets. \cite{buchhave2014} analyzed ground-based 
optical spectra of \textit{Kepler} Objects of Interest (KOIs) to measure spectroscopic metallicities and found three regimes of exoplanet sizes, split at $R_p$ $\sim 1.7~ R_{\oplus}$ and $R_p\sim 3.9~ R_{\oplus}$, distinguished by different (increasing with $R_p$) host star metallicities. Using the same data but with a more rigorous statistical analysis, \cite{schlaufman2015} instead favored a single, continuous relationship between planet radius and stellar metallicity. Interestingly, recent results show that (i) 2-6 $R_{\oplus}$ planets with orbital periods from 1-10 days ("hot Neptunes") show an increase in host star [Fe/H] compared to typical planet-hosting stars, similar to that of hot Jupiter ($R_p \geq 10 R_{\oplus}$) planets \citep{dong2017}, and (ii) at $<1$ day orbital periods, $\leq 2 R_{\oplus}$ planet host stars have significantly different metallicities than hot Jupiter host stars but similar metallicities as stars hosting $2-4 R_{\oplus}$ planets with 1-10 day periods \citep{winn2017}. These studies exemplify the (evidently) intricate relationship between the metallicities of host stars and the sizes and orbital configurations of the planets that form around them.

We intend to further characterize this intricate relationship by investigating how planet orbital period is tied to host star metallicity. This is a topic that has recently been explored by several other studies.

\cite{beauge&nesvorny2013} examined both confirmed exoplanet systems and \textit{Kepler} candidate multi-planet systems to show (i) a lack of small ($R_p \lesssim 4 ~R_{\oplus}$), short period ($P < 5$ days) planets around low metallicity (bulk [m/H] $< -0.2$ dex, from \citealt{buchhave2012}) stars, and (ii) a dearth of 4-8 $R_{\oplus}$ planets at $P \leq 100$ days around low metallicity stars. At the time, trends also held in the planetary mass versus period plane; e.g., M$_p ~sin~i < 0.05$ M$_{Jup}$ planets in short orbits were not found around [Fe/H] $< -0.2$ dex stars and planets between the masses of Neptune and Saturn with $P \leq 100$ days were mostly absent around [Fe/H] $<-0.2$ dex stars. The authors explained these observed trends with delayed formation and less planetary migration in metal-depleted protoplanetary disks. We note that the trends in \cite{beauge&nesvorny2013} are slightly reduced in significance when more up-to-date planet samples are considered (see \citealt{dawson2015}, discussed below).

Similarly, \cite{adibekyan2013} found from the
HARPS GTO radial velocity survey \citep{mayor2003, LoCurto2010,  santos2011} that $\sim$0.03 M$_{Jup}$ to 4 M$_{Jup}$ planets orbiting stars with [Fe/H] $<-0.1$ dex have longer periods than the same mass planets orbiting stars with [Fe/H] $> -0.1$ dex. Specifically, \citeauthor{adibekyan2012} find all M$_p ~ sin~i < 0.03$ M$_{Jup}$ planets orbiting [Fe/H] $> -0.1$ dex stars have periods $< 18$ days, and also 
suggest that smaller planets orbiting more metal-rich stars are more likely to migrate towards or form close to their host stars compared to planet orbiting more metal-poor stars. 

\cite{dawson2015} explored a theoretical framework motivated by these observational trends, combining analytical estimates for the formation of planetary embryos (that merge to form super-Earths and the cores of mini-Neptunes) with numerical simulations of atmospheric 
accretion in disks having varying solid surface densities. Interpreting their model predictions in the context of easily observed quantities (planet radius and host star metallicity), \citeauthor{dawson2015} find that disks with high solid surface density (metallicity) generate 2~$M_{\oplus}$ cores before the gas disk dissipates ($\sim$1 Myr), which enables the cores to more readily accrete significant atmospheres and thus increase their gas to rock fraction ($R_p$). Furthermore, these authors find a match with current observations -- i.e., that metal-rich stars lack rock-dominated ($R_p < 1.5$ $R_{\oplus}$) planets beyond $\sim 15$ day periods -- and suggest that this may indicate that embryo, and thus final core, masses grow faster at larger orbital distances in metal-rich versus metal-poor disks, thus producing gas-enveloped, larger $R_p$ planets.

Most recently, \cite{mulders2016} used over 20,000 stars observed by both \textit{Kepler} and LAMOST \citep{cui2012} to confirm that short period planets ($\lesssim 10$ days) are preferentially found around more metal-rich stars ([Fe/H] $\simeq 0.15 \pm 0.05$ dex), whereas longer period planets orbit roughly solar metallicity ([Fe/H]$\sim 0$) host stars. In the $P < 10$ day sample, it is the smallest radius planets ($< 1.7 ~R_{\oplus}$) that have the largest host [Fe/H] contrast compared to their similarly-sized but longer period counterparts, with an occurrence-weighted $\Delta$[Fe/H]~$\simeq~0.25 \pm 0.07$. \citeauthor{mulders2016}  suggest that their results may be evidence that the inner edges of protoplanetary disks around more metal-rich stars are closer in than around more metal-depleted stars. The trend observed by \citeauthor{mulders2016} is 
in contrast to the assessment by \cite{winn2017}, who comment that their metallicities (from \citealt{petigura2017}, using HIRES/Keck data from the California \textit{Kepler} Survey) of small planet host stars show no such period dependence. Differences in sample selection may influence the differences in the \citeauthor{mulders2016} vs. \citeauthor{winn2017} results. 

Many of the works above use moderate to high resolution optical spectroscopy to derive host star parameters. Indeed, the original \textit{Kepler} Input Catalog (KIC) was not intended for detailed studies of host star metallicity \citep{brown2011}, which motivated numerous follow-up spectroscopic campaigns to better characterize KOIs \citep[e.g.,][]{bruntt2012, buchhave2012, buchhave2014, everett2013, dong2014, brewer2016, petigura2017}. In this work we present a study of host star [Fe/H] versus planetary orbital period using high resolution near infrared spectroscopy
of KOIs taken by the Sloan Digital Sky Survey's Apache Point Observatory Galactic Evolution Experiment (APOGEE, \citealt{majewski2017}). In \S2 we discuss the APOGEE stellar parameter derivation, and validate the [Fe/H] and \teff values produced by APOGEE's automated stellar parameter pipeline (ASPCAP) by comparing its output to the results from several literature studies. In \S3 we explain the data collection for our KOI sample.
In \S4 we present our analysis of the KOI planet and host star parameters, focusing on orbital period and [Fe/H], and in \S5 and \S6 we discuss the interpretation of our results and final conclusions. 

%
%

\section{Validating APOGEE Spectroscopic Parameters}

All the data in this work were collected as part of APOGEE in the fourteenth Data Release (DR14, \citealt{dr14}) of the third and fourth Sloan Digital Sky Survey \citep{eisenstein2011, blanton2017}. APOGEE utilizes a multi-object spectrograph \citep{wilson2010, wilson2012} mounted on the Sloan 2.5\,m telescope \citep{gunn2006} to sample up to 300 sources simultaneously with high resolution ($R\sim 22,500$), high signal-to-noise ratio (SNR$>$100), $H$-Band (1.5--1.7\,\micron) spectroscopy. Details on the motivation and scope of the APOGEE survey are described in \cite{majewski2017} and the targeting is described in \cite{zasowski2013}. All of the data from APOGEE is processed through automated reduction and stellar parameter pipelines \citep{nidever2015, holtzman2015}, and the spectroscopic parameters used for the stars in our sample are derived from the Automated Stellar Parameters and Chemical Abundances Pipeline (ASPCAP). We give a brief overview of ASPCAP here for convenience, but for details on the pipeline we refer the reader to \cite{aspcap}. 

ASPCAP consists of two principal components: a \textsc{fortran90} optimization code (\textsc{ferre}, \citealt{allendeprieto2006}\footnote{Available from \url{github.com/callendeprieto/ferre}}) that compares the observed APOGEE spectra to synthetic libraries, and a multifunctional IDL wrapper used for bookkeeping and reading and preparing the input APOGEE spectra. \textsc{ferre} performs a $\chi^2$ minimization to find the best-fit set of atmospheric parameters (effective temperature, $T_\mathrm{eff}$; surface gravity, $\log g$; microturbulent velocity, $\xi_t$; and general solar-scaled metallicity, [M/H]) as well as C, N, and $\alpha$-element abundances from an interpolated library of synthetic ATLAS9 or MARCS model atmospheres. 
The atomic and molecular line list, gathered from the literature, has been updated regularly and is described most recently in \cite{shetrone2015} and Holtzman et al. (2017, in prep). 

Once fundamental atmospheric parameters are found, ASPCAP extracts individual chemical abundances by fitting spectral windows optimized for each element. 
Iron has dozens of Fe \textsc{i} lines in the $H$-band, and 
Fe abundances are computed using $\sim$55 spectral windows. ASPCAP provides both raw and calibrated values for all of its spectroscopic parameters. Calibrated \teff values are established using observations of globular and open clusters, and by requiring that there are no trends of abundances with $T_\mathrm{eff}$ in clusters (Holtzman et al. 2017, in prep).
In this study, we adopt the DR14 calibrated \teff and metallicity ([Fe/H]) values.

\begin{figure*}
   \centering
   \includegraphics[width=0.49\textwidth]{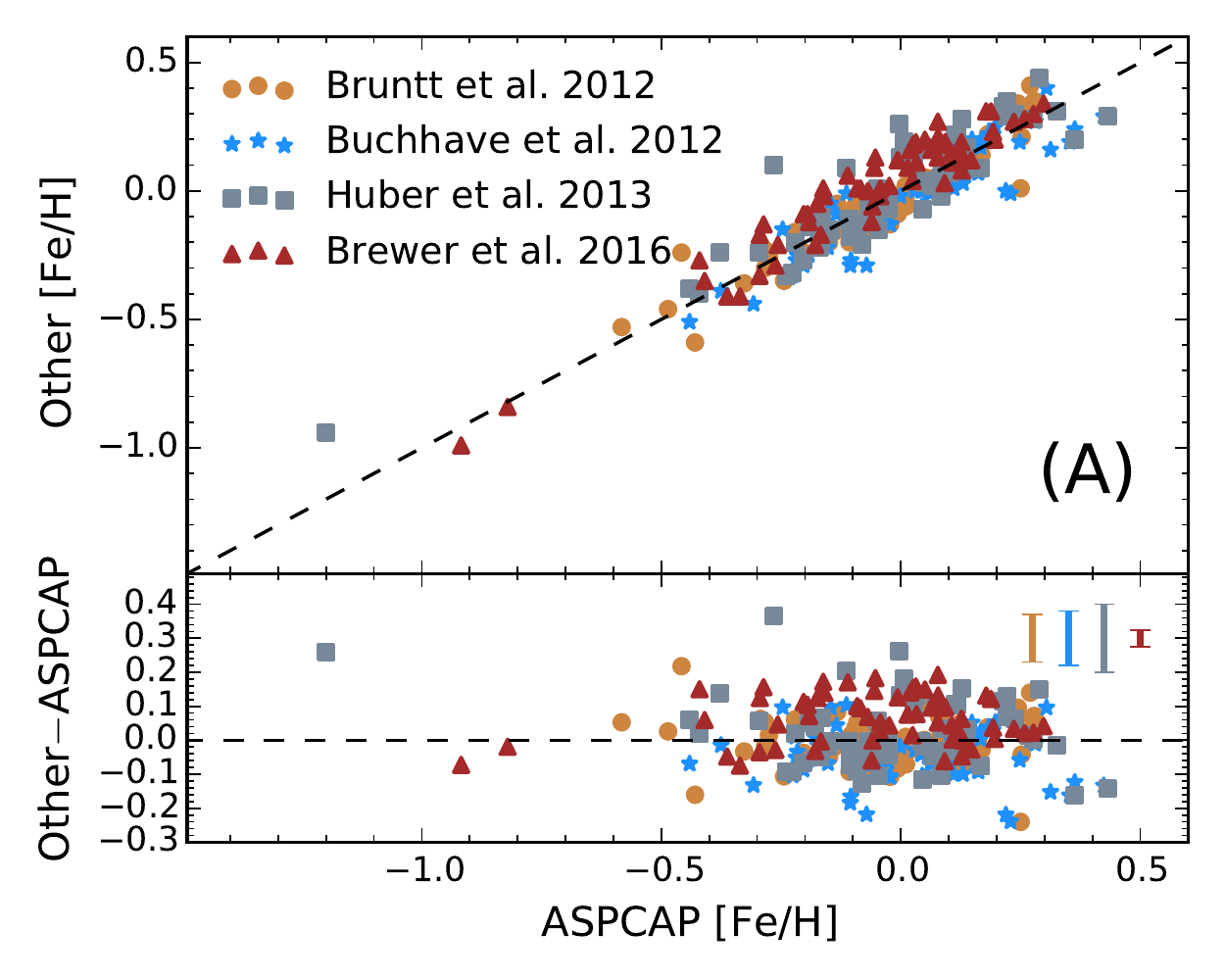} 
   \includegraphics[width=0.49\textwidth]{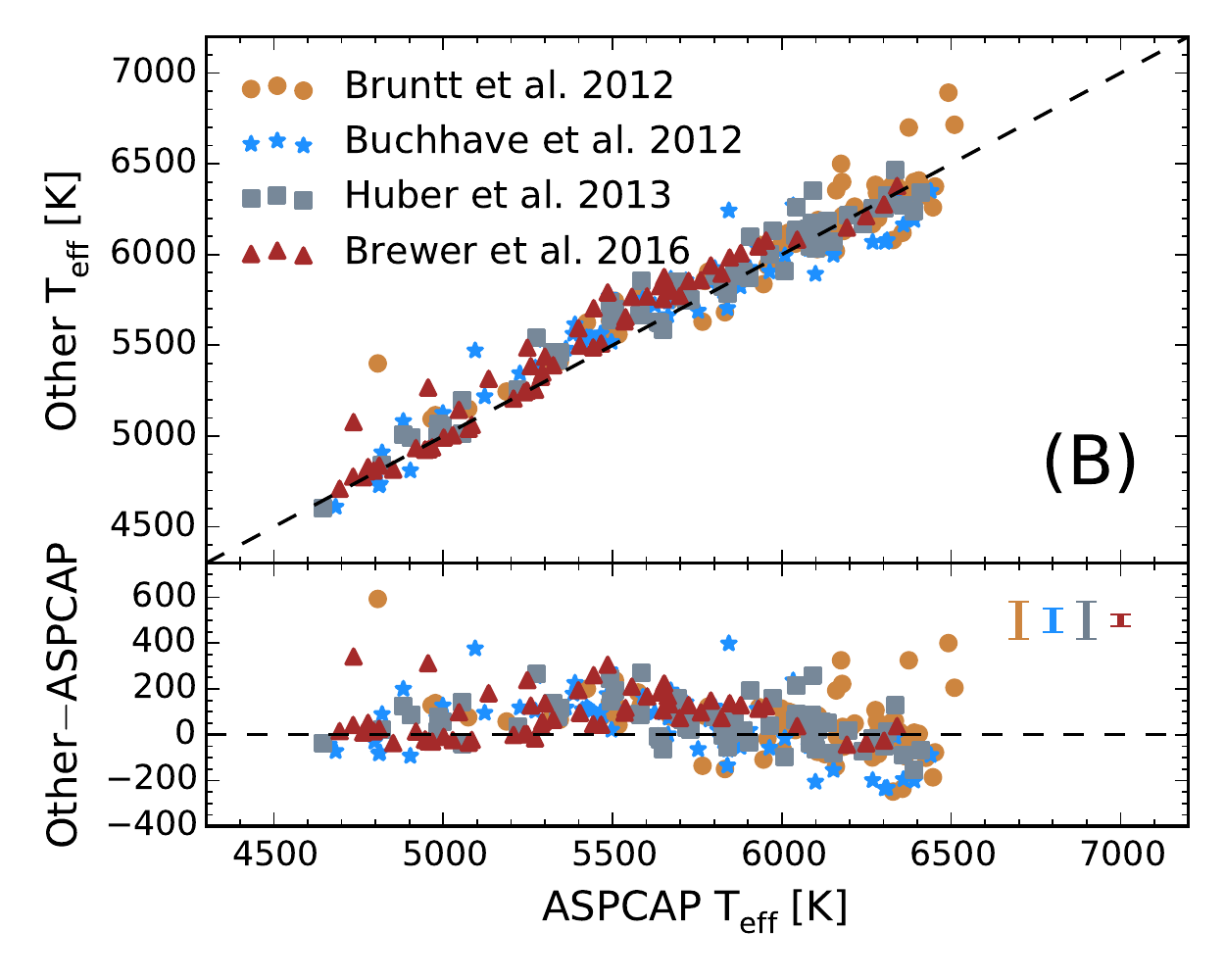} 
   \includegraphics[width=0.49\textwidth]{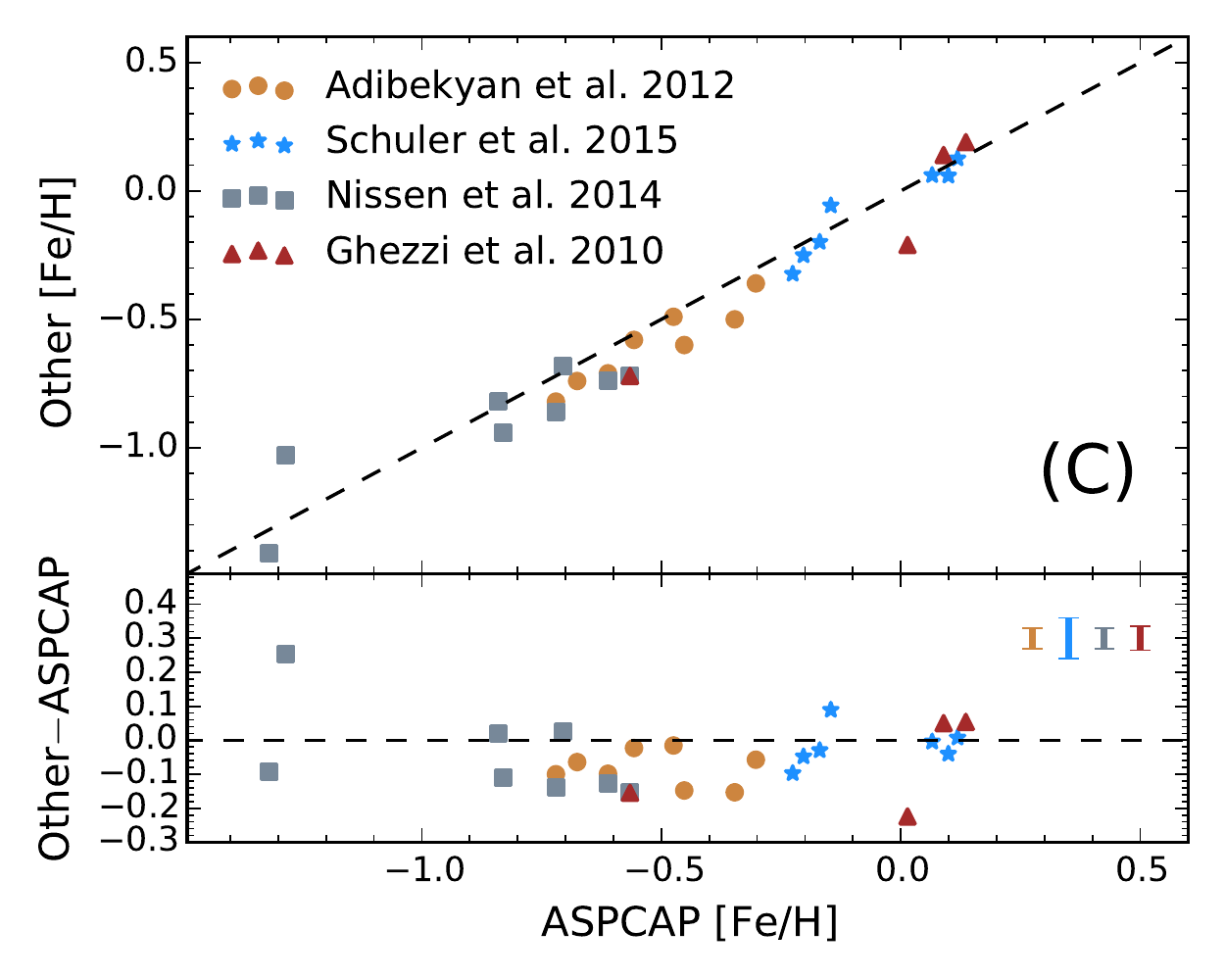}
   \includegraphics[width=0.49\textwidth]{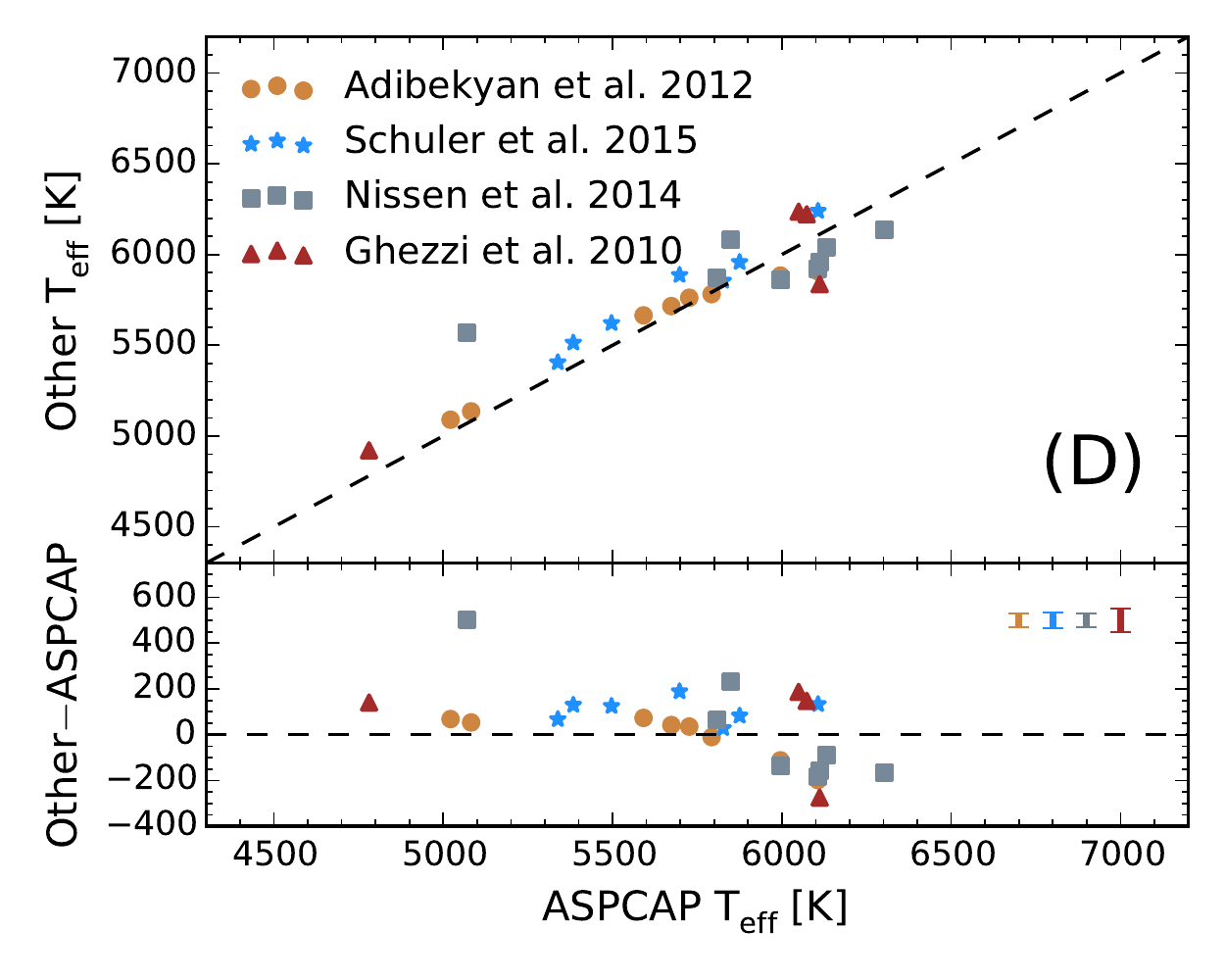}
      \label{fig:fig1}
   \caption{Comparisons of APOGEE [Fe/H] and \teff measurements to values determined in the literature. Each plot shows the \teff and [Fe/H] values from ASPCAP and the literature plotted against each other, as well as the difference (Other-ASPCAP). For the top panels, the dotted black line shows the one-to-one relationship, and on the bottom panels, the dotted black line shows the line of no difference. Typical errors from the literature studies are shown in the top right of each bottom panel. (A) ASPCAP's [Fe/H] determinations compared to measurements from synthesis studies. As a whole, the difference (Other$-$ASPCAP) shows good agreement with an RMS scatter of 0.09 dex, and a mean offset of 0.00 dex. (B) APOGEE \teff measurements compared to those from synthesis studies. Though there is a slight bump around \teff$\sim5500$~K, APOGEE shows excellent overall agreement with these surveys, with an RMS scatter of 129 K and mean offset of 57 K. (C) APOGEE's [Fe/H] determination compared to equivalent width studies in the literature. The difference in measured iron abundances by APOGEE and these studies show a mean difference of $-0.05$ dex and an RMS scatter of 0.11 dex. (D) APOGEE's \teff determination compared to equivalent width studies in the literature. The difference in measured temperatures by APOGEE and these dedicated studies show a mean offset of 36 K and an RMS scatter of 166 K. }
\end{figure*}

Because ASPCAP is optimized for red giants and not well tested for dwarfs (the topic of a separate publication in preparation), it is worthwhile to test the pipeline's performance against published spectroscopic studies of dwarfs. We select comparison studies focused on planet search targets, because the stellar samples are similar to those in our study (i.e., consisting mostly of FGK dwarfs). We compare first to four large surveys that derive stellar parameters using spectral synthesis \citep{bruntt2012, buchhave2012, huber2013, brewer2016}, 
each with enough stars to give a substantive comparison of ASPCAP's performance. These data and comparisons are shown in \hyperref[tab:surveys]{Table 1} and \hyperref[fig:fig1]{Figure 1, top panels}. We also compare ASPCAP's performance to equivalent width analysis studies of detailed chemical abundances \citep{ghezzi2010,adibekyan2012,nissen2014,schuler2015} to better gauge the pipeline's performance. These data and comparisons are shown in \hyperref[tab:studies]{Table 2} and \hyperref[fig:fig1]{Figure 1, bottom panels}. The equivalent width studies employ different methodologies, which allows us to compare ASPCAP against multiple strategies for deriving spectroscopic parameters. Summaries of these comparisons with ASPCAP's performance are given in \S2.2 and \S2.3. 

\subsection{ASPCAP Internal Errors on [Fe/H]}

Each of the comparisons described below represents an estimate of the relative accuracy of the ASPCAP stellar parameter results, but not their precision.  To first assess the internal error  on the ASPCAP [Fe/H] values, we turn to the sample of stars in the solar-age open cluster M67 that was observed by APOGEE, with parameters derived in the same way as our sample of KOIs.  From the  astrometric survey of \cite{yadav2008}, we selected the stars that were observed by APOGEE having the highest M67 membership probability ($\geq 90$\%), and also had ASPCAP uncalibrated $\log g$ values\footnote{The spectroscopic surface gravities for dwarfs in APOGEE DR14 are significantly lower than what is expected from stellar isochrones, and an acceptable calibration relation has not yet been developed by the ASPCAP team.} $\geq 4.0$ and measured [Fe/H] values. From this sample of 76 stars, the [Fe/H] median is -0.021 dex, the mean is -0.018 dex, and the standard deviation is 0.073 dex. However, as described below (\S4.1), in our analysis we include only high SNR ($>$100)
spectra and exclude stars with \teff$<4000$~K; performing the same cut to the M67 sample results in 46 stars with a [Fe/H] median of -0.016 dex, a mean of -0.011 dex, and a standard deviation of 0.053 dex. 

M67 is known to be chemically inhomogeneous -- \citet{liu2016} found a metallicity difference of $\sim0.05$ dex between a solar twin and a solar analog in M67, as well an enhancement of $\sim0.05$ dex in neutron-capture elements in the solar analog versus the solar twin. Additionally, using SDSS-III DR12 APOGEE data, \citet{bertrandelis2016} found the spread in [O/Fe] in cool, low-gravity stars (4000$<$\teff$<$4600 K, log $g<3.8$) in M67 to be higher ($\sigma$[O/Fe]$_{err} \sim 0.03$ dex) as compared to other solar or super-solar metallicity clusters NGC 6791 and NGC 6819, which show $\sigma$[O/Fe]$_{err} \lesssim 0.01$ dex. However, only a handful of dwarf stars in NGC 6791 and NGC 6819 were targeted by APOGEE, and even fewer make our SNR cut. Thus we adopt the $\sigma$[Fe/H] value from M67, 0.053 dex, as a conservative error (since a significant part of the spread in [Fe/H] is likely intrinsic to the cluster) on the ASPCAP metallicities for the KOIs in our analysis henceforth.

\subsection{Comparison to Literature -- Spectral Synthesis Studies}

\begin{deluxetable*}{ccccccccccc}
\tablecolumns{10}
\tablewidth{0pc}
\tabletypesize{\scriptsize}
\tablecaption{Spectral Synthesis Study Comparison Parameters \label{tab:surveys}}
\tablehead{ \colhead{APOGEE ID} & \multicolumn2c{ASPCAP}& \multicolumn2c{Bruntt et al. 2012} &\multicolumn2c{Bucchave et al. 2012} &\multicolumn2c{Huber et al. 2013} & \multicolumn2c{Brewer et al. 2016}  \\ 
 \colhead{ } & \colhead{\teff (K)} & \colhead{[Fe/H]} &\colhead{\teff (K)} & \colhead{[Fe/H]} & \colhead{\teff (K)} &
 \colhead{[m/H]} & \colhead{\teff (K)} & \colhead{[Fe/H]}  & \colhead{\teff (K)} &\colhead{[Fe/H]}  }

\startdata
2M19452396$+$4404359& 6425& 0.04& 6344 &0.01& -- & -- & -- & -- & -- & --\\
2M19455565$+$4400329& 4882& 0.08& -- &--& 5082& 0.03& 5009&-0.02&--&--\\
2M19480452$+$5024323& 5927& 0.10&--&--& 6040& 0.11&--&--&6008&0.18\\
2M19542140$+$4045024& 6096& -0.03&--&--&--&--& 6081& -0.03&--&--\\
2M20015142$+$4421140& 6070& -0.33& 6114& -0.36&--&--&--&--&--&--\\
\enddata
\tablecomments{This table is available in its entirety in a machine-readable form online. A portion is shown here for guidance regarding its form and content.}
\end{deluxetable*}

\cite{bruntt2012} utilized the analysis package \textsc{vwa} \citep{bruntt2010a} to derive stellar parameters and elemental abundances for a sample of 93 G dwarfs in the \textit{Kepler} field. Their data consist of high resolution $(R \approx 80,000)$, and high SNR $(\sim 200-300)$ optical spectra. The accuracy of their parameters resulted from adopting asteroseismic $\log g$'s, which they held fixed to derive the rest of their parameters. \cite{bruntt2012} reported typical uncertainties in their \teff and [Fe/H] determinations of $ 80~ \mathrm{K}$ and $ 0.07$ dex, respectively. The overlapping sample with APOGEE contains 71 stars. The difference (ASPCAP$-$Other) in the effective temperature ($\Delta T_\mathrm{eff}$) and iron abundance ($\Delta$[Fe/H]) determinations between ASPCAP and \cite{bruntt2012} have mean offsets and RMS scatter of $48\pm147$ K and $0.00\pm0.07$ dex.

\cite{buchhave2012} systematically derived spectroscopic and stellar parameters for a sample of 152 planet-hosting stars (PHS's) with 1500 observations from multiple telescopes and spectrographs, having SNR $\geq 30$. The final stellar parameters were determined from the average of all the measurements of each particular star. For this study, the authors developed their own analysis package, \textsc{spc}, designed to analyze spectra with low to modest SNR. The authors claimed typical uncertainties in the \teff and [Fe/H] of $50$ K and $0.08$ dex respectively, which were derived as the scatter among their sample of measurements for each star. The overlap between \cite{buchhave2012} and APOGEE is 75 stars. From these, the mean offset and RMS scatter for $\Delta$\teff and $\Delta$[Fe/H] are $43\pm147$ K and $-0.04\pm0.09$ dex, respectively. 

\cite{huber2013} produced a catalog of fundamental stellar parameters for 66 PHSs in the \textit{Kepler} field. \cite{huber2013} used a combination of \textsc{sme} \citep{sme} and \textsc{spc} to obtain initial guesses of the \teff, [Fe/H] and $\log g$. 
Stellar $\log g$'s were then fixed to asteroseismic solutions (derived using the initial \teff and [Fe/H] from \textsc{spc} and \textsc{sme}), and \cite{huber2013} re-derived \teff and [Fe/H] with the asteroseismic constraints in place. Like \cite{buchhave2012}, the \cite{huber2013} data come from multiple telescopes, have modest SNR, and were analyzed in the same way for their initial \teff and [Fe/H] measurements. The authors reported an average uncertainty of $82$ K and $0.1$ dex for \teff and [Fe/H], respectively. All 66 stars in 
\cite{huber2013} were observed with APOGEE. The mean offset and scatter for $\Delta T_\mathrm{eff}$ and $\Delta $[Fe/H] are $52 \pm 105$\,K and $0.02\pm0.10$ dex, respectively.  

\cite{brewer2016} used \textsc{sme} \citep{sme} to provide a catalog of spectral and stellar properties for 1615 FGK dwarfs. All observations were taken with the HIRES spectrograph \citep{vogt1994} at the Keck I telescope with resolution $R \approx 70,000$, but vary in SNR (about 25\% of their stars have SNR $<100$, and the rest have SNR$\geq 100$). \cite{brewer2016} note a strong dependence of their derived parameters with SNR. Of the \cite{brewer2016} subsample that was also observed with APOGEE, only five stars have SNR $<100$. These authors reported a typical uncertainty in \teff of $\sim 25 $ K and a typical [Fe/H] uncertainty between $\sim 0.01-0.04$ dex. Of the 1615 stars in their sample, 60 have also been observed by APOGEE. In this overlapping sample, we find a mean offset and scatter for $\Delta T_\mathrm{eff}$ and $\Delta $[Fe/H] of $82 \pm 126$ K and $0.06\pm0.10$ dex, respectively.

\subsection{Comparison to Literature -- Equivalent Width Studies}

\begin{deluxetable*}{ccccccccccc}
\tablecolumns{10}
\tablewidth{0pc}
\tabletypesize{\scriptsize}
\tablecaption{Equivalent Width Study Comparison Parameters \label{tab:studies}}
\tablehead{ \colhead{APOGEE ID} & \multicolumn2c{ASPCAP}&  \multicolumn2c{Adibekyan et al. 2012} &\multicolumn2c{Schuler et al. 2015} &\multicolumn2c{Nissen et al. 2014} & \multicolumn2c{Ghezzi et al. 2010}   \\ 
 \colhead{ } & \colhead{\teff (K)} & \colhead{[Fe/H]} &\colhead{\teff (K)} & \colhead{[Fe/H]} & \colhead{\teff (K)} &
 \colhead{[Fe/H]} & \colhead{\teff (K)} & \colhead{[Fe/H]}  & \colhead{\teff (K)} &\colhead{[Fe/H]}  }

\startdata
2M02360498$+$0653140& 4781& 0.01&--&--&--&--&--&--&4922&-0.21\\
2M02515835$+$1122119& 5809&-0.70&--&--&--&--&5873&-0.68&--&--\\
2M03402202$-$0313005& 5996& -0.72&5884& -0.82&--&--&5859&-0.86&--&--\\
2M13121982$+$1731016& 6112& -0.57&--&--&--&--&5956&-0.72&5837& -0.72\\
2M19134816$+$4014431& 5875& -0.17&--&--&5958&-0.20&--&--&--&-- \\
\enddata
\tablecomments{This table is available in its entirety in a machine-readable form online. A portion is shown here for guidance regarding its form and content.}
\end{deluxetable*}

In addition to the large surveys described above, we compare ASPCAP's results to a few select studies that have computed stellar parameters and derived elemental abundances through equivalent width (EW) measurements. Specifically, we compare ASPCAP's results with \cite{ghezzi2010}, \cite{adibekyan2012}, \cite{nissen2014}, and \cite{schuler2015}. In all of the following studies, stellar parameters were determined by adjusting the parameters until there was no correlation between the [Fe/H] values derived from  Fe \textsc{i} lines and the lower excitation potential ($\chi$) of the lines, nor between [Fe\textsc{ i}/H] and reduced EW [log (EW/$\lambda$)], and until there was agreement between abundances derived from Fe \textsc{i} and Fe \textsc{ii} lines. 

\cite{ghezzi2010} obtained spectra of a sample of 117 PHS's from the Fiber-fed Extended Range Optical Spectrograph (FEROS; \citealt{kaufer1999}). The setup chosen by \cite{ghezzi2010} resulted in spectral coverage from 3560 to 9200 \AA, and a nominal resolution of $R\approx 48,000$. The reported typical SNR per resolution element ranges from $\sim 200-500$. EWs were measured using the code \textsc{ares} \citep{sousa2007} and stellar parameters were derived using the 2002 version of \textsc{moog} \citep{sneden1973} assuming local thermodynamic equilibrium (LTE). The uncertainties for this sample were derived with the method outlined in \cite{gonzalez1998}. \cite{ghezzi2010} report typical \teff uncertainties ranging from $\sim30-70$ K and typical [Fe/H] uncertainties ranging from $\sim0.02-0.05$ dex. For the four stars observed by APOGEE and \cite{ghezzi2010}, the mean offset and RMS scatter in $\Delta T_\mathrm{eff}$ between ASPCAP and Ghezzi et al. is $51\pm195$ K, and the mean offset and scatter in $\Delta$[Fe/H] is $0.07\pm0.14$ dex, indicating a fair agreement.

\cite{adibekyan2012} obtained a sample of 1111 FGK dwarfs from the HARPS (High Accuracy Radial velocity Planet Searcher) GTO (Guaranteed Time Observations) planet search program \citep{mayor2003, LoCurto2010,  santos2011}. The spectra taken from the HARPS spectrograph \citep{mayor2003} have spectral resolution $R\approx 110,000$ and a SNR ranging from $\sim20-2000$, where 84\% of the sample has SNR $\geq$ 100. Their sample consists of dwarfs similar in \teff to the Sun, the majority of which lie within 4500K to 6500K and with metallicities ranging from -1.39 to 0.55 dex. The analysis was completed in a similar manner as \cite{ghezzi2010}, by assuming LTE, generating a grid of Kurucz model atmospheres \citep{kurucz1993}, and making use of \textsc{ares} and \textsc{moog}.
However, because \citeauthor{ghezzi2010} limited their line list to lines with $\log gf$ values measured in the lab, \cite{adibekyan2012}'s analysis benefits from a more extensive line list. 
The uncertainties from \cite{adibekyan2012} were determined by adding quadratically the uncertainties in the parameters of the atmospheric model and the scatter measured amongst the abundances of each individual line. Since HARPS is in the Southern Hemisphere, there are only eight stars that were also observed with APOGEE. The reported uncertainties were typically about 30K for \teff and 0.03 dex in [Fe/H]. The mean offset and rms scatter of $\Delta$\teff and $\Delta$[Fe/H] with respect to ASPCAP are $-6\pm92$ K and $-0.08\pm0.09$ dex, respectively.

\cite{nissen2014} measured the C/O ratio in a sample of 66 sun-like stars, with \teff ranging from 5400\,K to 6400\,K. The \cite{nissen2014} data are from both HARPS and FEROS, with the same configurations as \cite{adibekyan2012} and \cite{ghezzi2010}. The sample of stars observed with HARPS has SNR $\gtrsim 300$, while the FEROS-observed stars have a typical SNR $\sim 200$. \cite{nissen2014} derive \teff in their HARPS-FEROS sample using photometric calibrations, and initially derive [Fe/H] values by interpolating within the plane-parallel model atmosphere MARCS grid \citep{gustafsson2008}. They then derive more accurate [Fe/H] values by measuring the EWs of Fe \textsc{ii} lines using the Uppsala program
\textsc{eqwidth}. \citeauthor{nissen2014} report a typical uncertainty in [Fe/H] of 0.03 dex, and internal, one-sigma \teff errors of 30\,K. Though \citeauthor{nissen2014} do not explicitly describe how they derive their errors, they state their errors are drawn from uncertainties in their model atmosphere fits and equivalent width measurements. Comparing the eight stars in \citeauthor{nissen2014} that were also observed by APOGEE, the mean offset and rms scatter of $\Delta$\teff and $\Delta$[Fe/H] are $8\pm230$ K and $-0.04\pm0.14$ dex, respectively.

\cite{schuler2015} derived stellar parameters and elemental abundances for seven PHS's identified by \textit{Kepler}. The data were collected from HIRES on the Keck I telescope as part of the \textit{Kepler} Follow-up Observing Program (KFOP, \citealt{gautier2007}). The KFOP spectra have a spectral coverage of 3650-7950 \AA, and a spectral resolution of $R\approx 50,000$. \citeauthor{schuler2015} only considered data with SNR $\geq$ 150. Stellar parameters and abundances for this study were extracted using an LTE, curve-of-growth analysis. EWs were measured using the analysis package \textsc{spectre}, and abundances were derived from the 2014 version of \textsc{moog}, along with synthetic fits to the data interpolated from the ATLAS9 Kurucz model atmosphere grid. 
Uncertainties in \teff are reported as the difference between the adopted \teff value and the value that results in a $1\sigma$ correlation between the [Fe/H] vs. $\chi$ and reduced EW relations. \citeauthor{schuler2015} report uncertainties in \teff between 25\,K and 45K, and uncertainties in [Fe/H] between 0.04 and 0.08 dex. In the \citeauthor{schuler2015} sample, all seven stars in the Schuler et al. sample were also observed with APOGEE, resulting in a mean offset and scatter between the two studies for $\Delta$\teff and $\Delta$[Fe/H] of $110\pm119$ K and $-0.02\pm0.06$ dex, respectively.


Overall, after these various comparisons with multiple studies, we find that ASPCAP is accurate for \teff and [Fe/H] within the scatter, and agree with these multiple 
optical studies that utilize different methodologies.
We note that the mean offset in \teff (57\,K) versus the synthesis studies indicates that ASPCAP may be underreporting the \teff compared to these other studies. However, we find almost no offsets in \teff as compared to the studies of \cite{nissen2014} and \cite{adibekyan2012}.
Furthermore, because -57\,K is well within the RMS scatter for each of these studies, we do not consider it to be problematic for our purposes. Taking all these comparison studies into account, we find that the mean offset and RMS scatter in $\Delta$[Fe/H] (0.004 dex and 0.10 dex, respectively) are within the uncertainties required for this work. 

Having validated the performance of ASPCAP, we now move on to the study of a particular subset of APOGEE data consisting of of repeated observations of \textit{Kepler} objects of interest (KOIs) resulting in high SNR spectra.


%
%
\section{APOGEE KOI RV Sample}

The primary goal of APOGEE, now in its second phase APOGEE-2 \citep{majewski2016}, is to study the Milky Way through the radial velocities (RVs) and chemical abundances of as many as half a million stars, chosen to be primarily 
red giants across multiple stellar populations and Galactic regions. Additional science programs are also included in the survey, with one such program monitoring KOIs to search for false positives through RV variations \citep{Fleming2015}. APOGEE data reach an RV precision of $\sim$$100~\mathrm{m\,s^{-1}}$ \citep{troup2016}, allowing the search for eclipsing binaries and other grazing incidence geometries that may resemble transiting planets in the initial \textit{Kepler} reduction pipeline. The APOGEE survey will eventually observe $\sim$1050 KOIs with $\geq$18 epochs across five APOGEE-2 fields (roughly the size of \textit{Kepler} tiles), each KOI with a sufficient signal-to-noise ratio to get quality RVs at each epoch. As a result, the final "RV-normalized", \textit{summed} spectra over all epochs are of very high SNR (typically a few hundred), which allows for the derivation of high-precision stellar parameters and elemental abundances for planet-hosting stars. The APOGEE targets were chosen with the goal of observing all possible ``Confirmed" or ``Candidate" KOIs with $H<14$ in those five \textit{Kepler} tiles. Some KOIs were excluded from the sample on the basis of unphysical impact parameters and planet radii consistent with stellar values. Currently APOGEE has observed $\sim$600 KOIs from the Q1-Q16 catalog \citep{mullally2015}, orbiting $\sim$450 PHS's, each with between 10 and 28 epochs at the time of this study. 

Data concerning the orbital and planetary parameters for each KOI were gathered using the public NASA Exoplanet Archive, which provides the information in the form of interactive tables of confirmed and candidate planetary and stellar properties and includes a suite of integrated analysis tools \citep{akeson2017}. Use of this archive allows us to exclude known false positives and ensures that we are using the most up-to-date KOI dispositions in the literature. The specific targets included in our analysis, vetted from the $\sim$600 KOIs observed thus far by APOGEE-2, are described below (\S4.1).

%
%
\section{KOI Stellar Metallicity and Planet Period Relation}

\subsection{Selected Sample}

All of the stars in our KOI sample were observed as part of the APOGEE KOI Goal Program \citep{Fleming2015}, as described above. Initially, that consists of 624 KOIs and 450 PHS's. To ensure the quality of the data, we restrict our analysis sample using a series of APOGEE flags and other constraints.\footnote{Descriptions of the APOGEE flags can be found at \url{http://www.sdss.org/dr13/algorithms/bitmasks/\#APOGEE_TARGET1}} We first remove data with any of the \textsc{starflags}  \textsc{bad\_pixels, very\_bright\_neighbor}, and \textsc{low\_snr} set. It is worth mentioning that a fraction of our sample falls on the high-persistence region of APOGEE's "blue chip". However, we decide to keep these data since persistence effects were shown to be minimized in DR14 (Holtzmann et al. 2017, in prep). To exclude unreliable ASPCAP fits defined as values close to the edge of the model atmosphere grid,
we also remove data with any of the following \textsc{aspcapflags}: \textsc{teff\_bad}, \textsc{logg\_bad}, and \textsc{metals\_bad} set. Because the focus of this study is [Fe/H], 
we remove KOIs with the \textsc{paramflags} \textsc{gridedge\_bad}, \textsc{calrange\_bad}, \textsc{other\_bad}, and \textsc{param\_fixed} flags set, with respect to the [Fe/H] parameter. In addition, we require that all of the summed APOGEE spectra in our sample have SNR $>50$. 
Because the ASPCAP line list currently does not include FeH lines, which are important for modeling the metallicities of M dwarfs, we exclude all stars with  \teff$<4000$\,K (for a detailed discussion see \citealt{souto2017}).

In addition to the quality cuts described above, we also trim our sample based on the orbital periods and inferred planet radii of the KOIs. To ensure we avoid regions of parameter space associated with low survey completeness, we only analyze KOIs with orbital periods, $P<100$ days.\footnote{For an estimate of survey completeness as a function of planet radius and orbital period, see \cite{burke2015}}  For multiple planet systems, we only analyze the planet with the shortest orbital period.
We also restrict our sample to KOIs with inferred radii for the planet candidates $R_p < 20 ~ R_\oplus$. We anticipate any planet candidates having radii larger than this limit are likely to be eclipsing binaries (EBs). We correct for false positives from our sample by removing eight known EBs identified in the literature. We identify eight more likely binaries from visual inspection of spectra 
for which ASPCAP reported high $v\sin i$ values. In these cases, the reported $v \sin i$ values were a result of the combined spectra from the primary and companion stars.
To remove any more potential EBs from our sample, we also filter stars with high RV variability, which we define as the ratio of the scatter of the RV measurements to the error of the RV measurements, given by 
\begin{equation}
	\frac{\textsc{vscatter}}{\textsc{verr\_med}} > 17 ,
\end{equation}
where \textsc{verr\_med} is the median RV measurement error from all visits and \textsc{vscatter} is the RMS scatter of all the RV measurements. Because the RV measurement errors are often underreported in APOGEE \citep{troup2016}, we choose the cutoff as the median value for the 16 known binaries in our sample, which is $\sim$17. After these cuts, our final sample consists of 282 KOIs (all unique PHS's), listed in \hyperref[tab:kepler]{Table 3}.

The planet hosts in our sample are all FGK dwarfs with effective temperatures ranging from 4000\,K -- 6500\,K. The [Fe/H] of our sample range from -0.6 -- 0.4 dex. We note that this parameter space is well covered by our tests of ASPCAP using literature comparisons (\S2.2, \S2.3). The spectral SNR in our KOI sample from APOGEE have a wide range; 
the inner 68\% ranges in SNR from 70--280, while the asymmetric distribution peaks at SNR $\sim140$ with a tail to SNR $\gtrsim500$.

\begin{deluxetable*}{lccccccccccc}
\tablecolumns{12}
\tablewidth{0pc}
\tabletypesize{\scriptsize}
\tablecaption{Parameters of Selected Sample \label{tab:kepler}}
\tablehead{ \colhead{KOI} & \colhead{KIC} & \colhead{Period} &
  \multicolumn2c{Period Error}  & \colhead{Planet Radius} & \multicolumn2c{Planet Radius Error}  & \colhead{$K_p$} & \colhead{APOGEE ID} &\colhead{$T_\mathrm{eff}$}& \colhead{[Fe/H]}\\
 \colhead{ } & \colhead{} & \colhead{} & \colhead{Upper Bound} &
 \colhead{Lower Bound} & \colhead{} & \colhead{Upper Bound}  & \colhead{Lower Bound} & \colhead{} & \colhead{ } & \colhead{ }  \\
\colhead{ } & \colhead{} & \colhead{(days)} & \colhead{(days)} &
 \colhead{(days)} & \colhead{(R$_{\oplus}$)} & \colhead{(R$_{\oplus}$)}  & \colhead{(R$_{\oplus}$)} & \colhead{(mag)} & \colhead{ } & \colhead{(K)}& \colhead{(dex)}}

\startdata
K00041.02	&	6521045	&	6.89	&	2.28E-05	&	-2.28E-05	&	1.30	&	0.08	&	-0.07	&	11.20	& 2M19253263$+$4159249 &5827 & 0.07	\\
K00049.01	&	9527334	&	8.31	&	4.21E-05	&	-4.21E-05	&	2.74	&	0.43	&	-0.13	&	13.70	&	2M19285977$+$4609535& 5831 &  -0.07\\
K00084.01	&	2571238	&	9.29	&	3.83E-06	&	-3.83E-06	&	2.10	&	0.26	&	-0.10	&	11.90	&	2M19214099$+$3751064& 5437 & -0.02\\
K00100.01	&	4055765	&	9.97	&	1.22E-05	&	-1.22E-05	&	16.89	&	2.15	&	-4.29	&	12.60	&	2M19244270$+$3911581& 6336 &-0.32\\
K00103.01	&	2444412	&	14.91	&	1.28E-05	&	-1.28E-05	&	2.62	&	0.33	&	-0.17	&	12.59	&	2M19264400$+$3745057& 5485 & 0.06 \\
K00119.01	&	9471974	&	49.18	&	2.48E-05	&	-2.48E-05	&	8.20	&	0.50	&	-0.55	&	12.65	&	2M19381420$+$4603443& 5584& 0.33 \\
K00156.02	&	10925104&	5.19	&	8.92E-06	&	-8.92E-06	&	1.0	&	0.08	&	-0.07	&	13.74	&2M19362914$+$4820582& 4662 & 0.23 \\
\enddata
\tablecomments{This table is available in its entirety in a machine-readable form online. A portion is shown here for guidance regarding its form and content.}
\end{deluxetable*}

\subsection{Analysis \& Results}

As discussed in \S1, a correlation between an exoplanet's orbital period and the metallicity of its host star could indicate that protoplanetary disks with higher solid surface density cause planets to migrate or form closer to their host stars \citep{beauge&nesvorny2013,adibekyan2013,mulders2016}. Alternatively or in addition, such a correlation could mean metal-rich disks spawn planet cores faster at larger orbital distances, facilitating the cores growing into gas-enveloped planets with larger $R_p$ and thus causing an absence of strictly rocky planets at longer periods around metal-rich stars \citep{dawson2015}. Given the multiple interpretations and somewhat contradictory results regarding these host/planet properties \citep[e.g.,][]{winn2017}, we want to assess the presence and strength of the [Fe/H]-$P$ correlation within the APOGEE KOI sample.

To assess first the correlation between host [Fe/H] and orbital period, we perform two different non-parametric tests, calculating Kendall's rank correlation coefficient, $\tau$ \citep{kendall1938}, and Spearman's rank correlation coefficient, $\rho$ \citep{spearman}. Kendall's $\tau$ coefficient is $\tau = -0.21$, with a $p$-value, $p_\tau = 1.40\times10^{-7}$~(the equivalent of a $5.1 \sigma$ deviation from a normal distribution). Spearman's rank correlation coefficient is $\rho = -0.31$ with a $p$-value, $p_\rho = 1.67\times10^{-6} \, (5.1 \sigma)$. These results indicate that as the host star [Fe/H] increases, the orbital periods of the planets around those stars decrease. To verify the robustness of these correlations, we perform a Monte Carlo simulation with $10^5$ sets of data. For every simulated dataset, we add a perturbation which is randomly drawn from a normal distribution with our adopted $1\sigma$ uncertainty of 0.053 dex (see \S2.1), to the ASCAP-derived host [Fe/H] for each KOI. We then recalculate $\rho$ and $\tau$ for each simulated dataset. For Kendall's rank correlation, we recover the significance of this trend at a ${4.83^{+ 0.31}_{-0.30} \sigma}$ level, and for Spearman's rank correlation we recover a significance of  ${4.83^{+ 0.31}_{-0.31} \sigma}$, where the errors represent the inner 68\% of the posterior distribution.

\begin{figure*}
  \centering
   \includegraphics[width=0.49\textwidth]{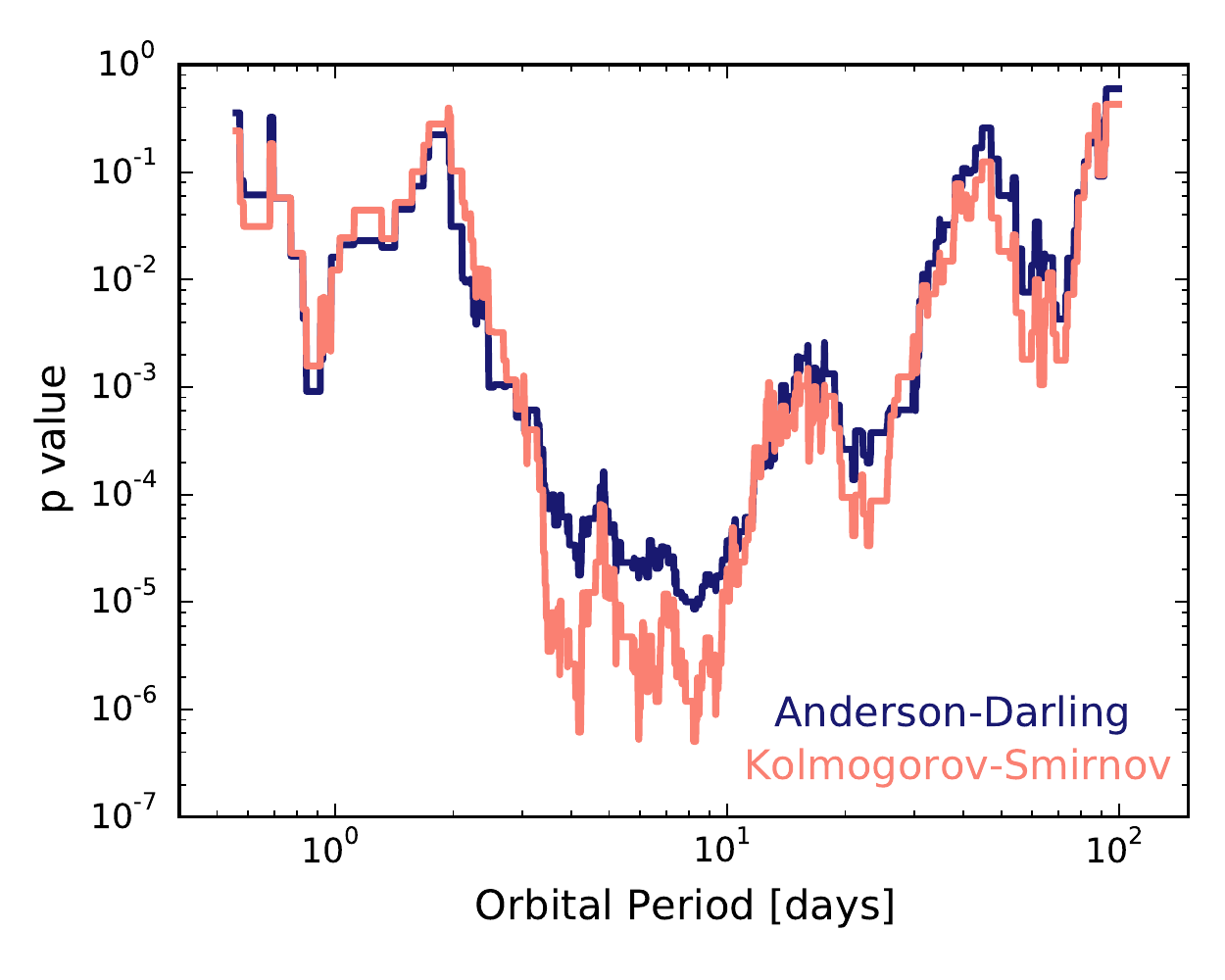} 
      \includegraphics[width=0.49\textwidth]{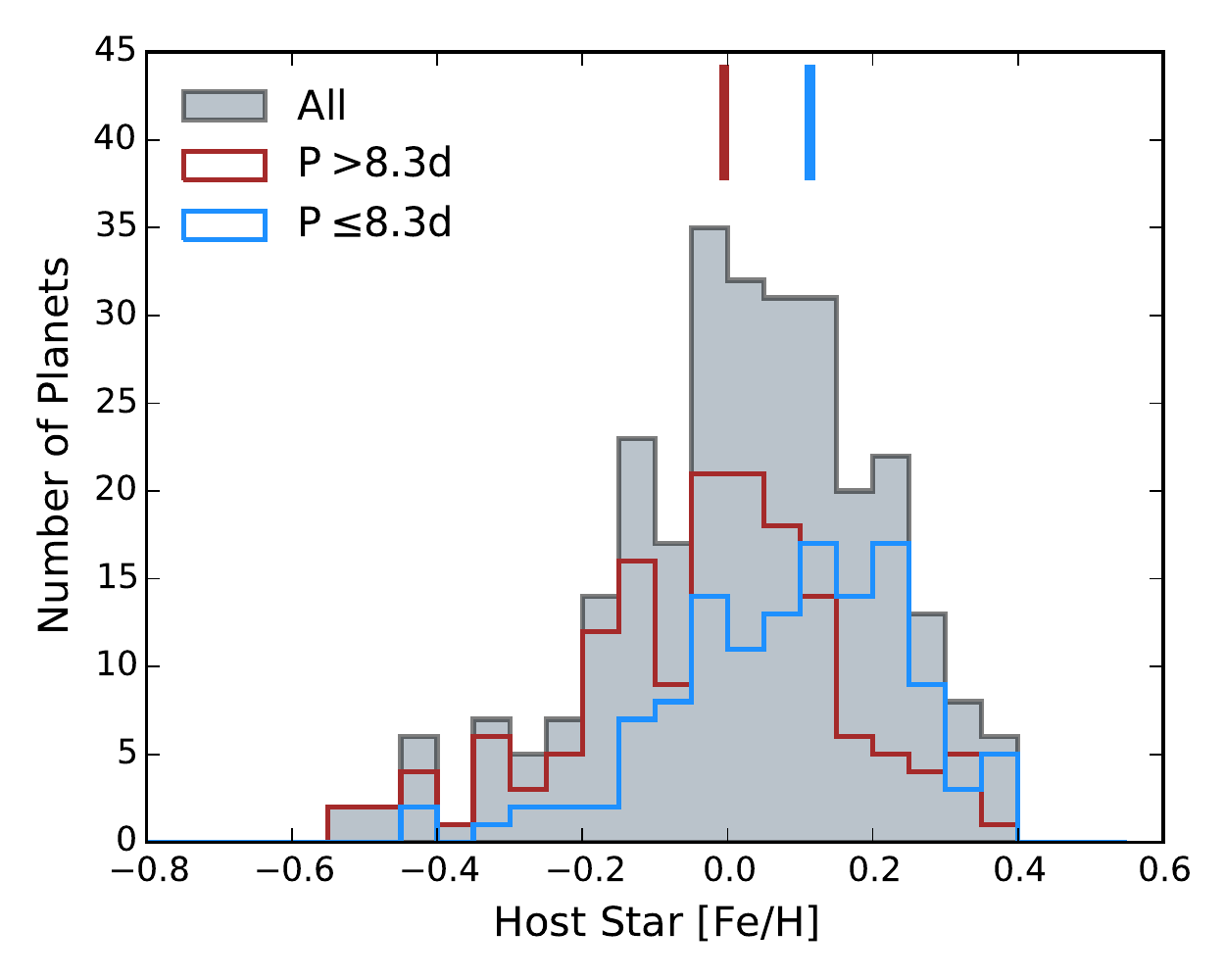} 
   \caption{(Left) The $p$ values of the Kolmogorov-Smirnov and Anderson Darling tests for the probability that the [Fe/H] distributions of exoplanet candidates above and below the given orbital period are drawn from the same parent distribution. {There is a statistically significant dip at ${P=8.2}$ days in our sample, found by both an Anderson-Darling test, and a Kolmogorov-Smirnov test}. (Right) Histogram of the host star metallicities of the long (red) and short (blue) period populations, split by \pcrit = {8.3} days. The combined distribution is shown in gray. The long period population peaks near solar metallicity while the short period population peaks above solar metallicity. The median host [Fe/H] is shown by the tick marks for the long (red) and short (blue) period populations.}
   \label{fig:fig2}
\end{figure*}

To analyze further the correlation between orbital period on host [Fe/H], we adopt a method similar to that employed by \cite{buchhave2014}. We generate $10^4$ test orbital periods equally separated in log space, spanning from the minimum to the maximum planetary orbital period in our sample. For each of these periods, $P_i$, we divide our KOI sample into two bins, one where the KOIs have orbital period $P > P_i$ and one with $P \leq P_i$. We then use a two-sample Kolmogorov-Smirnov (KS) test, as well as a $k$-Sample Anderson Darling (AD) test for redundancy, to determine the likelihood that the host star iron abundances from the ``short'' versus ``long'' period bins are drawn randomly from the same parent distribution. We find a critical period, \pcrit, where this likelihood is minimized (see \hyperref[fig:fig2]{Figure 2}). In the case where the minimum $p$ value is equal among more than one period, it is because we are oversampling our period distribution. In this case, we take the mean. {Within our dataset we find \pcrit~= 8.3 days by the KS test and the AD test, with $p$ values of ${p_{ks} = 5.0\times10^{-7} \, (4.9 \sigma)}$ and ${p_{ad} = 8.6\times10^{-6}\, (4.3\sigma)}$.} To test the robustness of this critical period, we perform a Monte Carlo analysis and simulate $10^4$ sets of data, resampling as we did above, using the typical [Fe/H] uncertainty of $0.053$ dex and the $P$ uncertainties as reported by the NASA Exoplanet Archive, which have a median of $4\times10^{-5}$ days. {We recover the significance with both the KS test and AD test at the ${4.5^{+0.4}_{-0.4}\sigma}$ level. We find \pcrit $={8.3^{+0.1}_{-4.1}}$ days for both the KS test and AD test, which is consistent with our original findings.} This method thus discovers two unique [Fe/H] populations within our dataset, one that is super-solar on average and contains planets orbiting closer to the host star, and one that is solar metallicity on average and contains planets that orbit farther from their host. That is, planets with orbital periods ${P\leq 8.3}$ days have statistically more metal-enriched hosts than planets with ${P>8.3}$ days. This is the main finding of this paper. We note that performing the same tests on host star [Fe/H] and planetary semi-major axis ($a$), instead of period, produces similar results, with $a_{\rm{crit}}= {0.07}$ AU.

In \hyperref[fig:fig2]{Figure 2}, we also note two other, less significant dips at $P\sim 22$ days and at ${P\sim70}$ days. The AD test found the shorter period dip at $P\sim 21$ days and the KS test found it at $P\sim 23$ days. For the longer period dip, the minimum $p$-value found with the KS test is at ${P=63}$~days, while the minimum found with the AD test is at ${P=71}$~days. Running a Monte Carlo analysis with $10^4$ data sets, similar to the above but restricting our analysis within the range $15~\mathrm{days} <P<45~\mathrm{days}$, {we find the period that minimizes the $p$-value to be ${P = 22.7^{+0.4}_{-1.5}}$~days for both the AD and KS tests. However, the significance with which we recover this period is only ${3.5^{+0.5}_{-0.5}\sigma}$ for both the AD and KS tests. Because of the lowered significance,} we are not comfortable claiming significance at this period within our study. However, there is theoretical motivation to support a critical period around 23 days (see \S5.2), and it warrants further work. To test the third dip, we perform the same analysis restricting our test periods to $P>45$ days, but the results are inherently less trustworthy considering the uneven sample sizes of KOIs with $P<65$~days and $P>65$~days. The period that minimizes the $p$-value for the AD and KS tests in this longest period range is ${69^{+4}_{-9}}$~days, with a significance of ${2.8_{-0.4}^{+0.4} \sigma}$ for each test. For the same reasons as the $P\sim23$~day dip, we do not consider this a significant minimum, and {presently favor a two-population model split only at \pcrit${=8.3^{+0.1}_{-4.1}}$~days. }

The iron abundance for the short period population ($P\leq$\pcrit) is super solar, with a median [Fe/H] of {0.11} dex and a {standard deviation of 0.17 dex.} The long period population ($P>$\pcrit) is consistent with solar metallicity and has a {standard deviation of 0.18 dex.} To test whether the means of these two samples differ significantly, we perform a Mann-Whitney U-test \citep{mann1947}  between these two populations. The Mann-Whitney U-test gives the probability that two separate populations have the same underlying mean. The test returns a $p$ value of ${p_{mw} = 2.28\times10^{-7}\,(5.0\sigma)}$ that the short and long period populations have the same mean metallicity. Thus, we can safely reject the null hypothesis that these distributions have the same mean, which is consistent with the results of the Kolmogorov-Smirnov and Anderson-Darling tests that the two [Fe/H] populations, separated by \pcrit, are sampled from different parent populations.  

\begin{figure}
   \centering
   \includegraphics[width=0.49\textwidth]{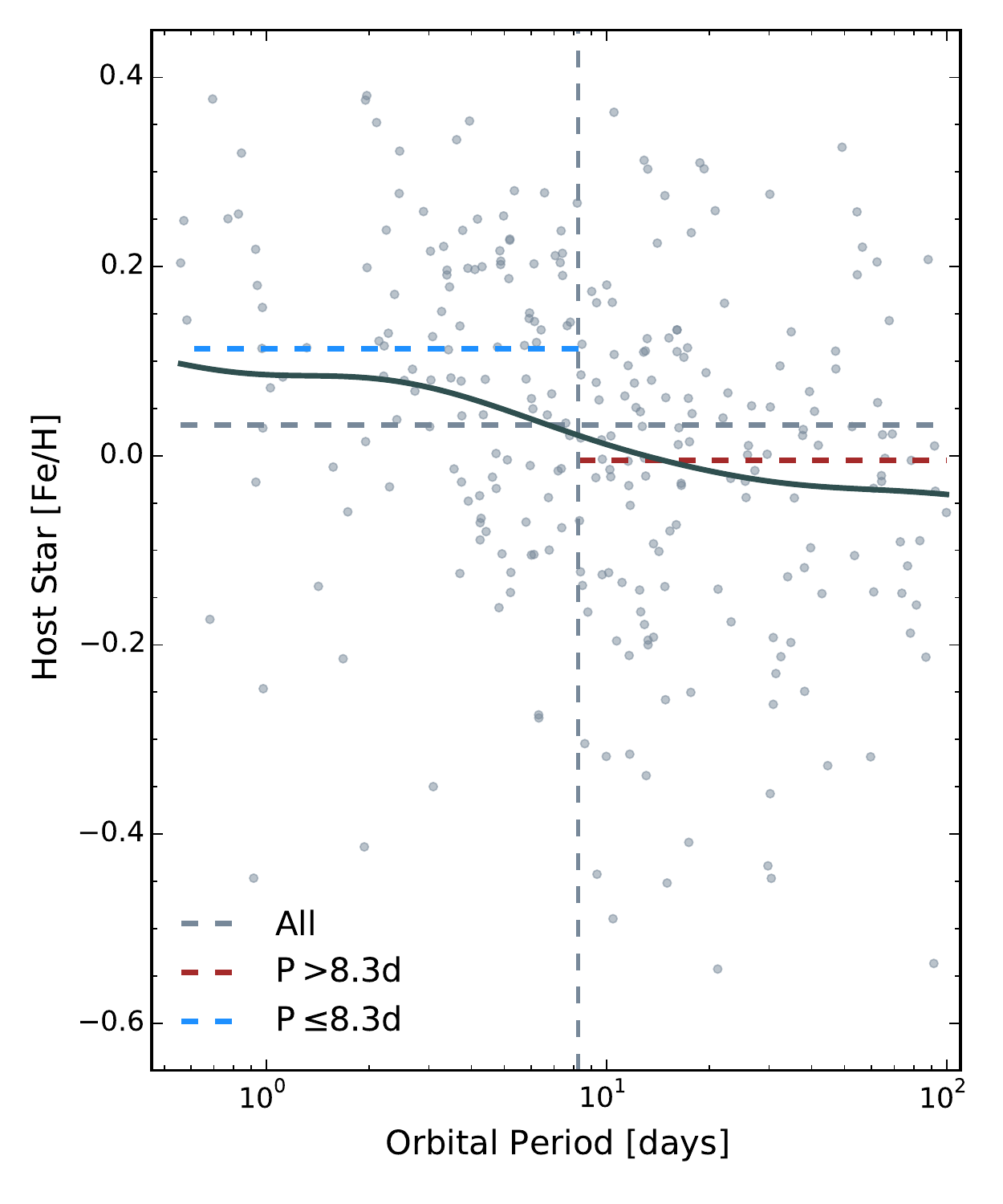} 
   \caption{Host star metallicity as a function of exoplanet orbital period. The gray points are the KOIs in our sample. The vertical dashed line is at $P={8.3}$ days, which separates our short period and long period populations. The horizontal dashed lines show the median of the short period (blue), long period (red), and combined (gray) populations. The combined population has a median of 0.03 dex. The median metallicity of the short period population is super solar at {0.11} dex, while the median of the long period population is 0.00 dex. The solid gray line is the kernel regression of the mean metallicity. The kernel regression shows a steady decrease from a maximum of $\sim {0.10}$ dex at the shortest periods, to a minimum of $\sim {-0.04}$ dex at the longest periods.}
   \label{fig:fig3}
\end{figure}

To further analyze the significance of the correlation in our sample, we follow the analysis of \cite{mulders2016} and use the Nadaraya-Watson estimator \citep{nadaraya1964,watson1964} to calculate how the mean iron abundance varies with orbital period (see \hyperref[fig:fig3]{Figure 3}). The kernel regression of the mean metallicity, $\overline{[\mathrm{Fe/H}]}_\mathrm{KOI}$, as a function of orbital period is given by 
\begin{equation}
	\overline{\mathrm{[Fe/H]}}_\mathrm{KOI}(P) = \frac{ \sum^{n}_{i=0} [\mathrm{Fe/H}]_i  \, K(\log(P/P_i), \sigma)}{\sum^{n}_{i=0}   K(\log(P/P_i), \sigma)}, 
\end{equation}
where the $n$ is the number of exoplanet candidates in the sample, $ [\mathrm{Fe/H}]_i$ and $P_i$ are the host star metallicity and orbital period of each exoplanet candidate, respectively, and we use a log-normal kernel with constant bandwidth, $\sigma$, given by
\begin{equation}
	K(\log x, \sigma) = \frac{1}{\sqrt{2\pi}\,\sigma} e^{-0.5 (\log x/\sigma)^2} \;,
\end{equation} 
where $x$ is an arbitrary, dimensionless variable. In line with \cite{mulders2016}, we adopt a bandwidth of $\sigma=0.29$.

The kernel regression of the mean metallicity is plotted over all the KOIs in our sample in \href{fig:fig3}{Figure 3}, along with the median [Fe/H] of the combined, short period, and long period samples. Using the kernel regression as a proxy for the mean [Fe/H], {we find that the maximum mean metallicity ([Fe/H] $\sim0.10$ dex) occurs at the shortest period in our sample, and the minimum mean metallicity in this trend ($\sim -0.04$ dex) occurs at the longest period in our sample.} {This difference of 0.14 dex is larger than the difference in the median metallicity between the long and short period sample by $\sim0.03$ dex.} However, the kernel regression clearly shows a decrease in mean [Fe/H] in our sample at even the longest orbital periods. 

%
%
\section{Discussion}

\subsection{Metallicity-Period Correlation}

Our results are consistent with those of \cite{mulders2016}, who find in their sample an increase of $0.15 \pm 0.05$ dex in host star metallicity for exoplanets orbiting at or interior to 10 days, as compared to longer-period planets. We find a statistically significant break in the KOI host star [Fe/H]-period distribution in our sample {at 8.3 days}, with shorter period planets orbiting stars with a {median [Fe/H] of 0.11$\pm$0.17 dex and longer period planets orbiting stars with a median [Fe/H] of 0.00$\pm$0.18 dex. The Mann-Whitney and Kolmogorov-Smirnov probabilities in our data are ${5.0\sigma}$ and ${4.9\sigma}$, respectively,} compared to \cite{mulders2016} whose Mann-Whitney and Kolmogorov-Smirnov probabilities were comparable at $4.6\sigma$ and $4.3\sigma$, respectively. While our sample is significantly less than half the size ({282} candidate or confirmed planets versus Mulders' 665), our data are measured from higher resolution, high SNR spectra, producing internal errors on [Fe/H] of only $\sim 0.053$ dex and typical offsets from literature values of $\sim 0.00 \pm 0.09$ dex. The \cite{mulders2016} study used LAMOST [Fe/H] values, measured from $R \sim 2000$ optical spectra, typically with SNR $\leq 100$, and typical [Fe/H] internal errors (evaluated by way of repeat observations of some stars) of $\sim$0.055 dex and typical literatures offsets (using their calibrated [Fe/H] values) of $\sim -0.06 \pm 0.18$ dex. Thus, our study showcases an advantage APOGEE has over other surveys of similar scale. Even with our significantly smaller sample size compared to the \cite{mulders2016} sample, we are able to recover the same correlation with {greater} confidence. 

\subsubsection{Required Precision to Find the Trend}

Given our smaller sample size, what is the minimum [Fe/H] precision required to be able to find the trend with orbital period that we do? To determine this precision, we replace the assumed [Fe/H] error (0.053 dex) with larger and larger errors until the resulting uncertainties on the significance of the Kolmogorov-Smirnov and Anderson Darling tests drop below $\sim 3 \sigma$. With our well-vetted (e.g., removal of stars showing signs of binarity in their radial velocity variations) sample of high resolution, high SNR data, the greatest value that the mean [Fe/H] error can take on is $\sim$0.1 dex to still recover the observed trend with at least $3\sigma$~significance. 

\subsubsection{Possible Mechanisms}

\begin{figure}
\centering
	\includegraphics[width=0.49\textwidth]{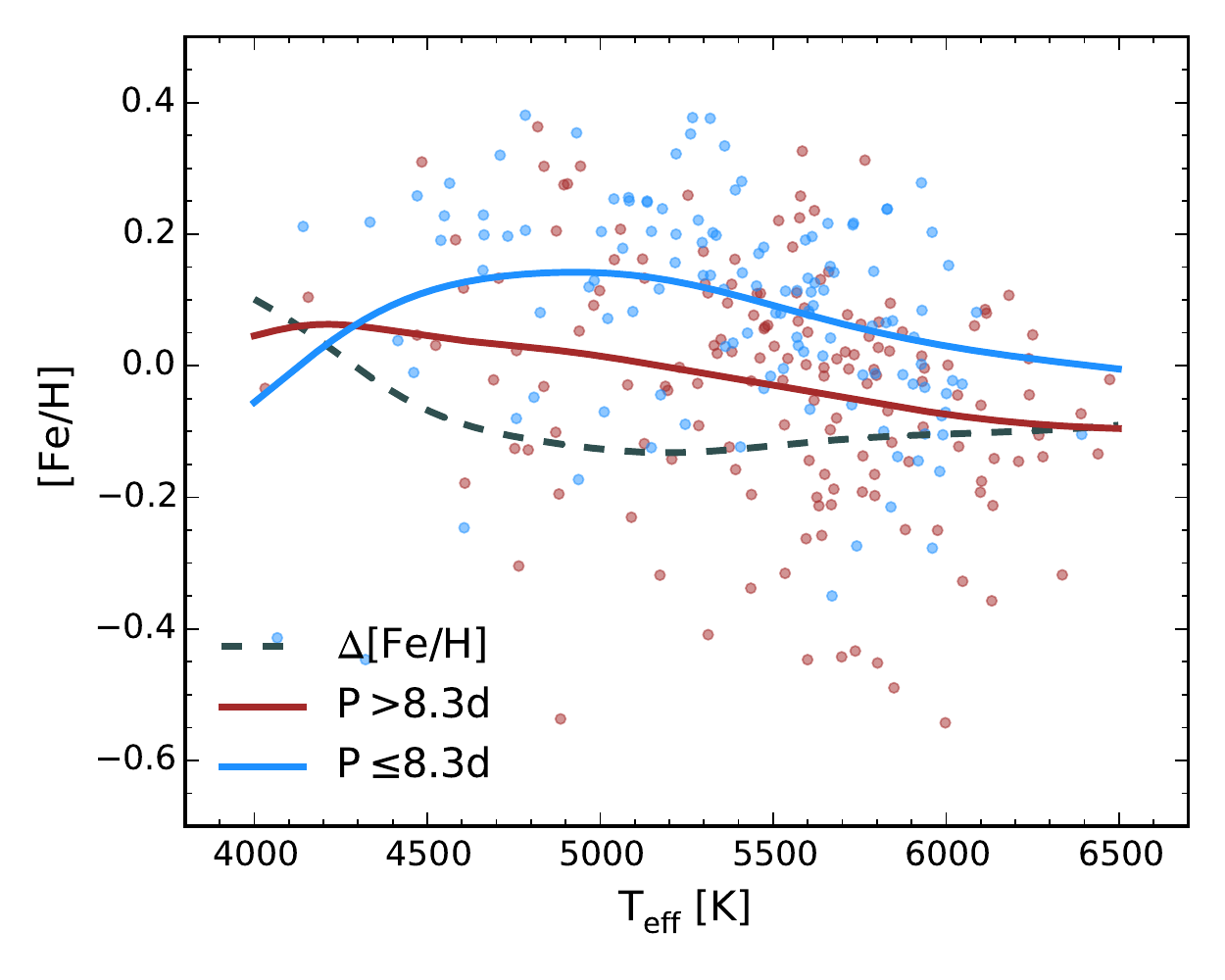}
    \includegraphics[width=0.49\textwidth]{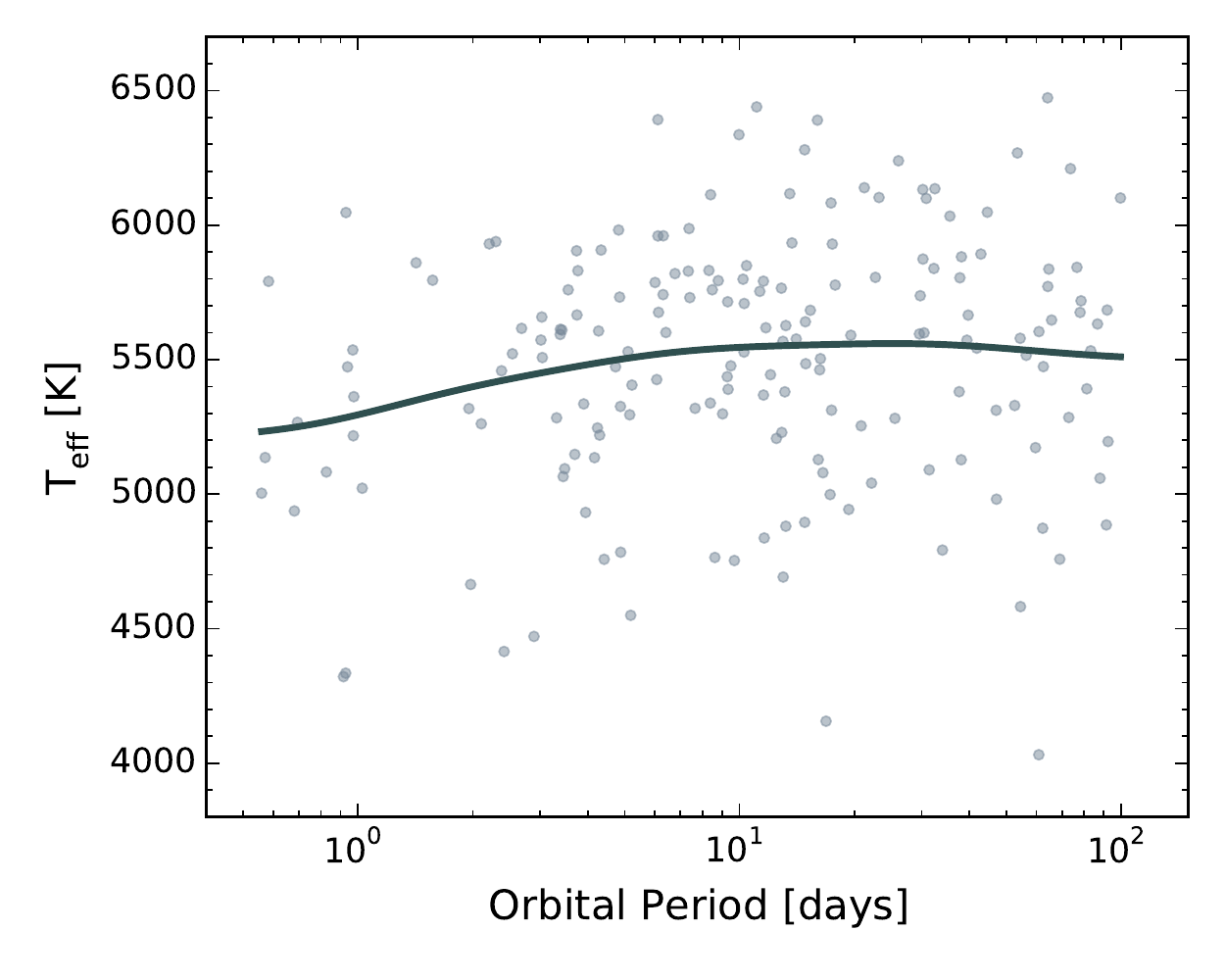}
   \caption{(Top) Kernel regression of the mean [Fe/H] as a function of PHS \teff for the short-period (blue) and long-period (red) populations, with the difference of the two plotted as a gray dashed line. The upturn at cool temperatures is most likely a result of low-number statistics. Hotter than $\sim$4500 K, the two distributions show a constant offset of $\sim${0.10} dex, indicating  no difference in host star \teff between the two samples. (Bottom) Stellar \teff versus planetary orbital period. The gray line shows the kernel regression of the mean \teff. If the orbital period-[Fe/H] correlation is related to the dust sublimation radius, we would expect to see a statistically significant positive correlation in this plot, which we do not. Hence there is no evidence for a dust-sublimation effect.}
   \label{fig:fig4}
\end{figure}

What is the physical mechanism responsible for the observed trend between planetary orbital period and host star [Fe/H]? One possible explanation is that the dust sublimation radius (between 0.05 and 0.1 AU, or $\sim 4-12$ days around a solar mass star -- \citealt{muzerolle2003}, \citealt{eisner2005}, \citealt{pinte2008}, \citealt{min2011}) in protoplanetary disks correlates with host star metallicity, and that it represents a semi-major axis cutoff inward of which no solids contribute to forming planets. If this were the case, we would expect a correlation between host star \teff (as the dust sublimation radius is known to depend on the stellar luminosity and/or stellar mass), and orbital period. To test this correlation, we performed Kendall's $\tau$ and Spearman's $\rho$ correlation tests {among our KOI sample. Because we are testing a potential semi-major axis cutoff, we include only the closest planets in the case of multiple-planet systems, as we do with the rest of this study}. We find $\tau = 0.10$ with $p_\tau={0.05 \; (1.7\sigma)}$ and $\rho=0.14$ with $p_\rho={0.06 \; (1.6\sigma)}$, which indicates that there is no significant correlation between \teff and orbital period. To test further  whether there is a difference in the host \teff among the short-period and long-period subsamples, we performed a KS and AD test, similar to the above analysis. {For the KS test, we calculate $p_{ks} = {0.2 ~ (0.8\sigma)}$ and for the AD test we calculate $p_{ad} = {0.07 ~ (1.3\sigma)}$}. Thus, both of these tests indicate that there is no significant difference in the temperature distributions of the host stars in the short-period and long-period samples. \hyperref[fig:fig4]{Figure 4}, left, shows that while there is a slight decrease in [Fe/H] from cooler to hotter \teff, both the shorter period/metal-rich and longer period/solar-metallicity subsamples decrease together with a roughly constant metallicity offset of {$\sim$0.10 dex.} Because the correlation between \teff and [Fe/H] is the same for both subsamples, and the shorter period and longer period subsamples do not have significantly different temperature distributions (\hyperref[fig:fig4]{Figure 4}, right), we do not believe this decrease in [Fe/H] at higher \teff affects our conclusions. Thus, our data rule out dust sublimation effects as a possible explanation for the metallicity-orbital period correlation. \cite{mulders2015} also ruled out the hypothesis of the dust sublimation radius controlling the semi-major axis cutoff in planetary occurrence rates, although these authors were testing a dependence of semi-major axis on stellar mass, not metallicity.

	A second possible explanation for the observed trend between planetary orbital period and host star [Fe/H] is that planets around higher metallicity stars migrate inward and are ``trapped'' closer to their host stars \citep[e.g.,][]{kuchner&lecar2002,rein2012,plavchan&bilinksi2013}. This closer location could be related to the dust sublimation radius -- for which we do not see evidence for as a factor in creating the P-[Fe/H] trend as described above -- or the co-rotation radius, where gas is accreted onto the stellar surface. Usually the co-rotation radius is within the dust sublimation radius (\citealt{najita2007} and references therein), and the former also depends on the angular velocity of the star, which does not have a strong stellar mass dependence during the pre-main sequence stage (\citealt{bouvier2013} and references therein). Thus it is difficult to assess the likelihood of this ``planet trapping'' explanation within the context of this work. However, as discussed below, small, short period planets around more metal-rich stars are generally rock-dominated, indicating accretion and migration mainly \textit{within} the snow line. Thus their migration distance could not have been very far. The snow line is at a few AU around a solar-mass star after $\sim$1 Myr, but this distance for any given disk will depend on the stellar mass, and also parameters like the heating mechanism(s), dust grain opacities, and the timing of planet formation \citep[e.g.,][]{kennedy&kenyon2008,martin&livio2012,mulders2015b,xiao2017}.

Finally, we consider the possibility that the correlation between planet orbital period and stellar metallicity could be a byproduct of rocky planet ingestion by the star driven by inward migration \citep[e.g.,][]{melendez2009,ramirez2010,ramirez2014,schuler2015,mack2014,liu2014,teske2015,teske2016a,teske2016b,bedell2017}. For example, \citet{mack2014} compared the elemental abundances of the twin stars in the planet-hosting wide binary system HD~20782/81. Both stars in that system host close-in planets within $\sim$0.2~AU, and \citet{mack2014} found that both stars are significantly enriched in refractory elements. Those authors further used a model of planet ingestion to show that the enhanced refractory abundances were consistent with a scenario in which the observed close-in planet pushed 10--20~$M_{\oplus}$ of rocky material (perhaps in the form of small rocky planetessimals) into the surface convection zone of the star. Presumably, this is a consequence of the inward migration of the observed close-in planet. Thus one possible explanation for the correlation we report here is that the planets in our sample that managed to migrate very close to their host stars ($P < {8.3}$~d) were also more likely to shepherd other small rocky planets into the star, thus elevating its surface metallicity.

In theory, a testable prediction of this rocky material ingestion scenario would be a pattern of increasing stellar elemental abundances with elemental condensation temperature ($T_c$), as found by \citet{mack2014} (and, originally in solar twins, \citealt{melendez2009}). However, we are currently unable to perform this test due to the uncertainties associated with dwarf stars abundances derived by ASPCAP in DR14; as noted in \S2, ASPCAP is optimized for red giants and not well tested for dwarfs.\footnote{We did attempt to make a preliminary assessment of the differences between the slopes of $T_c$-abundance trends for the short versus long period samples within our 300 stars with planets, but observed no difference between the slope distributions. If the short period planet host stars had ingested more refractory material, we would expect them to show more positive slopes.} Moreover, while giant planet inward migration has been suggested as a mechanism for ``pushing'' refractory material onto the host star, we know of no studies suggesting the inward migration of small planets would be capable of creating a similar abundance signature in their host stars. 

Interestingly, the scenario envisioned above could also potentially provide a natural explanation for the specific orbital period {(8.3~d)} dividing the metal-rich versus metal-poor samples. Previous work on the rotational evolution of pre--main-sequence stars has suggested a bimodal distribution of stellar rotation periods, the break between slow and rapid rotators occurring at $\sim$8~d \citep[see, e.g.,][and references therein]{Choi_Herbst:1996}. The two groups of rotators have been interpreted by some authors as the result of different angular momentum histories, possibly due to rotational braking by those stars whose protoplanetary disks survive longer, draining angular momentum from the star via magnetic connection between star and disk at the co-rotation radius as noted above. If inward migration of planets, coupled with ingestion of rocky material by the star, is also related to the rapid dispersal of the protoplanetary disk, the result could be a natural division at $\sim$8~d of those stars that were more likely to ingest refractory elements (i.e., show elevated surface metallicity), and more likely to host a close-in planet with an orbital period shorter than $\sim$8~d. 

If shorter period planets usually orbit more rapidly rotating stars, based on our observed planet $P$-[Fe/H]$_{star}$ trend we would predict that higher metallicity stars should show faster rotation and shorter disk lifetimes, on average. However, testing this prediction is complicated by the tendency for all stars to slow their rotation on Gyr timescales, regardless of their early rotational histories; we cannot know a priori the original rotation periods of the stars in our sample. To check whether their present-day stellar rotation periods show any trend with stellar metallicity or planetary orbital period, we cross-matched our sample with the \citet{McQuillan2013} and \citet{Walkowicz&Basri2013} catalogs of \textit{Kepler} stellar rotation periods, resulting in 82 stars in common. In this subsample we see no trends between stellar rotation period and stellar metallicity or planet orbital period. Moreover, near-infrared observations of young stars in clusters of various metallicites find that lower metallicity stars have \textit{shorter} protoplanetary disk lifetimes \citep{Yasui2009,Yasui2010}, and these observations are supported by models \citep{Ercolano&Clarke2010} and simulations \citep{Nakatani2017} of protoplanetary disk evolution. This contradicts the prediction that shorter period planets orbiting more metal rich stars can be explained by shorter disk lifetimes (and by extension faster stellar rotation periods). Additionally, the idea of bimodal rotation rates among young stars has itself been disputed \citep[see, e.g.,][]{stassun1999}. 

Thus we conclude that, while it is still plausible that the main trend we have discovered here could be the result of ingestion by stars of rocky material due to the inward migration of planets, it does not appear to be a consequence of star-disk interaction in the context of stellar rotational evolution.

\subsection{Planet Radius as a Third Dimension of the Correlation}
\begin{figure}
   \centering
   \includegraphics[width=0.5\textwidth]{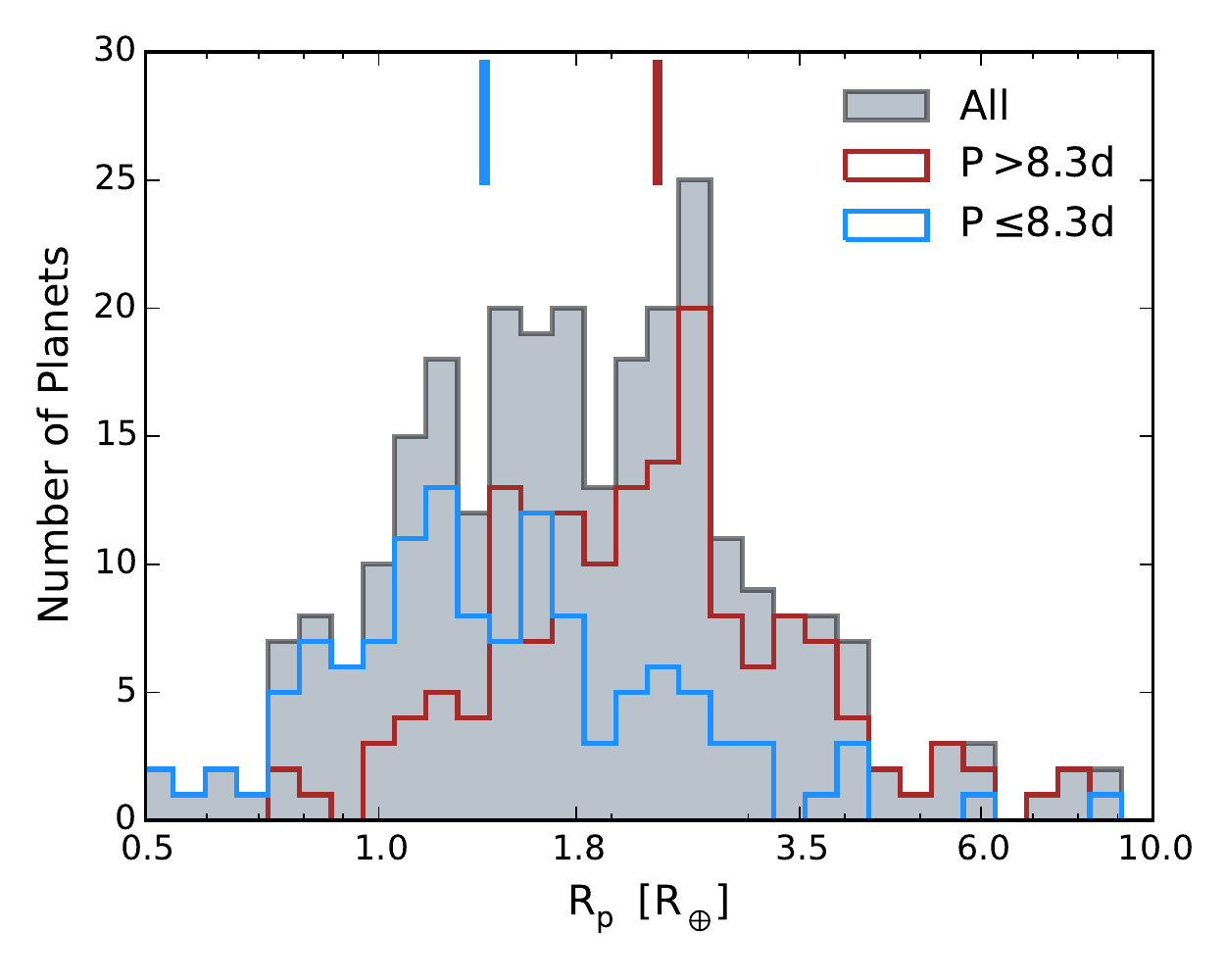} 
   \caption{Histogram of KOI radii of the long (red) and short (blue) period population, split by {\pcrit = 8.3 days.} The combined distribution is shown in gray. These are the same colors corresponding to the metal-enriched (blue) and solar-metallicity (red) distributions in \hyperref[fig:fig2]{Figure 2}, right panel. The tick marks at the top denote the median planet radius of the short-period (blue) and long period (red) populations.}
   \label{fig:fig5}
\end{figure}

Specifically, \citet{mulders2016} find an occurrence-corrected $\Delta$[Fe/H] between planets interior and exterior to 10 days that varies with planetary radius, from 
$\Delta$[Fe/H]$= 0.25 \pm 0.07$ dex for $R_p < 1.7~ R_{\oplus}$ planets to $0.08 \pm 0.05$ dex for $1.7~ R_{\oplus} \leq R_p < 3.9 ~R_{\oplus}$ planets to $0.10 \pm 0.12$ dex for $R_p \geq 3.9 ~ R_{\oplus}$ planets. We also find that the short period (metal-rich host star) planets in our sample are statistically smaller (${ p_{mw} = 7.1\sigma, p_{ks} = 6.7\sigma }$) than the planets at longer periods (around less metal-rich stars). The median, mean, and standard deviation of our metal-rich/short period planet population are {1.37, 2.01, and 1.91 $R_{\oplus}$, respectively, versus 2.29, 2.74, and 1.91 $R_{\oplus}$, respectively}, in our solar-metallicity/long period planet population; typical errors on $R_p$ are 0.02 $R_{\oplus}$. Interestingly, the break in [Fe/H]-period space in our sample also appears to coincide with the reported ``radius gap'' around 1.8 $R_{\oplus}$ defined by \cite{fulton2017} (see \hyperref[fig:fig5]{Figure 5}). While metal-poor dwarf stars will generally have larger radii than metal-rich dwarf stars, potentially influencing any trends with $R_p$ and host star [Fe/H] \citep{gaidos&mann2013}, the metallicity bias (the difference between [Fe/H] values of underlying population of stars versus those around which transiting planets are detected) for the stars in our sample is $\lesssim 0.02$ dex, below our measured precision. We will further explore this break in the $R_p$-[Fe/H]-orbital period distribution within the APOGEE Kepler sample in an upcoming publication. 

Recently, \citet{owen&wu2017} constructed a relatively simple analytical model, building on their previous numerical models \citep{owen&wu2013}, of a low-mass planet -- core with a gas-envelope -- and how it changes with time under the influence of its host star flux. As their previous results (and those of \citealt{lopez&fortney2013} also showed), the radius distribution observed by \citet{fulton2017} is matched by an ``evaporation valley'', such that smaller radius planets represent the bare cores of planets that have had their H/He envelopes photo-evaporated within the first 100 Myrs. \citet{owen&wu2017} show specifically that the two peaks in $R_p$ observed by \citeauthor{fulton2017} arise because the timescale for mass-loss is longest when the planet's radius has doubled in size due to an accreted volatile envelope of approximately a few percent the total planet mass. Their model is only able to reproduce the observed results when the planet envelopes are composed of primoridal H/He, \textit{not} water, and when the cores are roughly Earth-like in composition ($\rho \sim 5.5$ g~cm$^{-3}$). In summary, \cite{owen&wu2017} show that the small, short period \textit{Kepler} planets likely form from one parent population, with one average composition, and have a bimodal period-radius distribution due to envelope evaporation.

Almost simultaneous with \citet{owen&wu2017}, \citet{jin&mordasini2017} produced work independent of \citet{owen&wu2017} that compared theoretical models of planet formation, thermodynamical evolution, and atmospheric escape of rocky-cored versus icy-cored planets to the results of \cite{fulton2017}. Similarly, \citet{jin&mordasini2017}'s goal was to understand better, in a statistical sense, how evaporation depends on planetary bulk composition. \cite{jin&mordasini2017} suggest -- assuming the radius gap is due to atmospheric evaporation -- that small, short period planets have mostly rocky cores made of silicates and iron, not mostly icy cores made of frozen H$_2$O, CO$_2$, CH$_4$ and/or NH$_3$. Since planets with mostly icy cores can only form \textit{beyond} the snow line, this indicates that close-in low mass planets accreted mainly \textit{within} the snow line, even considering migration (that is, migration must have been within the snow line). Classifying observed planets based on their ice mass fractions (derived from the mass and radius and an internal structure model, \citealt{mordasini2012b}), as well as the planet's $R_p$ and semi-major axis $a$, the authors find eight categories of planets (see their Figures 6 and 7) that exhibit a clear compositional gradient with increasing planet radius. Interestingly, six of the eight planet categories requiring a rock-dominated composition ($R_p \lesssim 1.6$~$R_{\oplus}$) are found within $\sim$0.09 AU ($\sim$10 days around a solar mass star), and all eight of the rock-dominated planets are found within $\sim$0.17 AU ($\sim$26 days around a solar mass star). We see a second, less significant dip at $P\sim23$ days in \hyperref[fig:fig2]{Figure 2}, perhaps corresponding to this second rocky planet orbital period limit.  

Putting this all together, the following picture emerges: Most short period, small planets have rocky cores, but their size, volatile content, and thus density is (in a general sense) tied to their orbital period, and thus by this study, their host star metallicity. The trend we see predicts that planets with little to no volatiles, and thus the smallest $R_p$~values, should be in the closest orbits and thus around more metal-rich host stars. \cite{jin&mordasini2017} caution that they do not see convincing evidence of a positive volatile content gradient with increasing semi-major axis, but that this is expected for evaporation. We note that Kolmogorov-Smirnov and Anderson Darling tests do not reveal a significant difference between the orbital periods of \citealt{jin&mordasini2017}'s 25 Type 1 and 3 planets versus their seven Type 6 planets (from their Table 1), but that with a larger number of small planets with well-constrained masses and radii, we can better test our prediction. Our results, as interpreted in the context of \citeauthor{jin&mordasini2017}'s theoretical framework, suggest that in the hunt for small, rock-dominated planets with little to no gaseous envelopes, one should be looking around more metal-rich stars.

%
%
\section{Conclusions} 

In this work, we aim to characterize the intricate relationship between host star metallicity and planet orbital period, as it relates to the context of planet formation. We first demonstrate the veracity of ASPCAP's metallicities by comparing a sample of 221 FGK dwarfs in the APOGEE survey that also have quality parameters via optical spectroscopy in the literature. Then, using a sample of {282} short period ($P<100$~days) \textit{Kepler} exoplanets and exoplanet candidates observed by the APOGEE KOI goal program and the associated parameters derived from ASPCAP, we have characterized the correlation between planet orbital period and host star metallicity. In particular, we've found the following:

\begin{itemize}

\item There is a statistically significant correlation between host star [Fe/H] and planetary orbital period that is characterized by a critical period, \pcrit = ${8.3^{+0.1}_{-4.1}}$ days, below which planets preferentially orbit more metal-rich stars. This corresponds to a semi-major axis of {$\sim$0.07} AU for a solar mass star and is consistent with the drop in occurrence rate at $\sim$0.1 AU found by \cite{mulders2015}. 

\item The minimum precision in [Fe/H] needed to see this trend within our carefully vetted sample is $\sim$0.1 dex. While this correlation has been seen in other studies (e.g. \citealt{mulders2016}), the precision in APOGEE's abundance determinations allows us to find this correlation with higher confidence levels in a significantly smaller sample than what has been used for other planet host surveys of similar scale. 

\item Planets in the short-period/high-metallicity population have significantly smaller radii than the long period population (${p_{mw} = 7.1\sigma}$). Based on previous work on the ``evaporation valley'', this suggests that the population of planets around more metal-rich stars is mostly rocky and lacking substantial atmospheres, while the population of planets around more metal-poor stars have thicker atmospheres. Thus, to optimize the number of close-in, rocky exoplanets discovered around FGK dwarfs, transit surveys should prioritize super-solar metallicity stars. 

\item Based on the results of \cite{jin&mordasini2017}, we suspect that the critical period of {8.3} days may be tied to the bulk composition of the exoplanet population in a statistical sense. In addition to this period, we find some evidence for a second, less convincing critical period at $P\sim23$ days, which may also correlate with the exoplanet population's composition. Although we do not currently believe this period is significant, its agreement with the results of \cite{jin&mordasini2017} is intriguing enough to warrant further investigation. 

\item We hypothesize that there is some protoplanetary disk inner-radius with a metallicity-dependence at the time of planet formation that allows small, rocky planets to either form or migrate closer in to their host star in metal-rich conditions. Such an inner radius may be the dust-sublimation radius, but we would expect this radius and thus orbital period to correlate strongly with the host \teff, and see no such correlation. The inner radius may instead be the gas co-rotation radius, but with our given observations it is hard to 
 assess the likelihood of this explanation. Alternatively, the period-metallicity correlation that we observe may be the result of rocky planet ingestion, driven by inward planet migration. In this scenario, planets migrate inward and in the process shepherd rocky material (perhaps in the form of planetesimals) onto their host star, resulting in an increased surface metallicity. At this time the precision of APOGEE dwarf star abundances across a range of condensation temperatures preclude a robust test of this hypothesis.
\end{itemize}

APOGEE provides a valuable resource for characterizing exoplanet host stars from the \textit{Kepler} mission. In particular, studies of planetary architecture coupled with accurate metallicities, as presented here, can provide new constraints on planet formation that could not otherwise be obtained from smaller, more focused studies. In addition, APOGEE measures [Fe/H] to a level of precision that stands out from other spectroscopic surveys of similar scale, positioning APOGEE in a valuable area for future work in characterizing exoplanets and their host stars. 


\acknowledgments
We would like to thank the anonymous referee, whose insightful comments improved the quality of our work and helped to find a bug in our code.

RFW, SRM, NT, and MFS were partially supported by the National Science Foundation under grant No. AST-1616636.  VS and KC acknowledge that their work here is supported, in part, by the National Aeronautics and Space Administration under Grant 16-XRP16\_2-0004, issued through the Astrophysics Division of the Science Mission Directorate. DAGH was funded by the Ram\'on y Cajal fellowship number RYC-2013-14182 and DAGH and OZ acknowledge support provided by the
Spanish Ministry of Economy and Competitiveness (MINECO) under grant AYA-2014-58082-P.

This research has made use of the NASA Exoplanet Archive, which is operated by the California Institute of Technology, under contract with the National Aeronautics and Space Administration under the Exoplanet Exploration Program. This research has also made use of the Tool for OPerations on Catalogues And Tables \citep{topcat}.

Funding for the Sloan Digital Sky Survey IV has been provided by the Alfred P. Sloan Foundation, the U.S. Department of Energy Office of Science, and the Participating Institutions. SDSS-IV acknowledges support and resources from the Center for High-Performance Computing at the University of Utah. The SDSS web site is \url{www.sdss.org}. SDSS-IV is managed by the Astrophysical Research Consortium for the Participating Institutions of the SDSS Collaboration including the Brazilian Participation Group, the Carnegie Institution for Science, Carnegie Mellon University, the Chilean Participation Group, the French Participation Group, Harvard- Smithsonian Center for Astrophysics, Instituto de Astrof\'{i}sica de Canarias, The Johns Hopkins University, Kavli Institute for the Physics and Mathematics of the Universe (IPMU) / University of Tokyo, Lawrence Berkeley National Laboratory, Leibniz Institut f\"{u}r Astrophysik Potsdam (AIP), Max-Planck-Institut f\"{u}r Astronomie (MPIA Heidelberg), Max-Planck-Institut f\"{u}r Astrophysik (MPA Garching), Max-Planck-Institut f\"{u}r Extraterrestrische Physik (MPE), National Astronomical Observatory of China, New Mexico State University, New York University, University of Notre Dame, Observat\'{a}rio Nacional / MCTI, The Ohio State University, Pennsylvania State University, Shanghai Astronomical Observatory, United Kingdom Participation Group, Universidad Nacional Aut\'{o}noma de M\'{e}xico, University of Arizona, University of Colorado Boulder, University of Oxford, University of Portsmouth, University of Utah, University of Virginia, University of Washington, University of Wisconsin, Vanderbilt University, and Yale University.

\facilities{Sloan, \textit{Kepler}} 
\software{{Numpy \citep{numpy}, FERRE \citep{allendeprieto2006}, ASPCAP \citep{aspcap}}}

\bibliographystyle{yahapj}
\bibliography{references}


\end{document}